\documentclass[11pt,a4paper]{article}

\usepackage{jheppub_kim}
\usepackage{rotating}
\usepackage{graphicx,epsfig}
\usepackage{amsmath}
\usepackage {amssymb}
\usepackage{subfigure}

\usepackage{relsize}

\def\beq{\begin{equation}}
\def\eeq{\end{equation}}
\def\bea{\begin{eqnarray}}
\def\eea{\end{eqnarray}}

\begin{document}
\begin{flushright}
FU-PCG-15
\end{flushright}

\title{ Current density and  
conductivity through modified gravity in the graphene with defects}

\author[a,b]{Alireza Sepehri}

\author[c,d]{Richard Pincak}

\author[e]{Kazuharu Bamba}

\author[f,g,h,i]{Salvatore Capozziello }

\author[j,k,l]{Emmanuel N. Saridakis}

 \affiliation[a]{Faculty of Physics, Shahid Bahonar University, P.O. Box 76175, Kerman,
Iran}

 \affiliation[b]{Research Institute for Astronomy and Astrophysics of
Maragha (RIAAM), P.O. Box 55134-441, Maragha, Iran }

\affiliation[c]{Institute of Experimental Physics, Slovak Academy of Sciences,
Watsonova 47,043 53 Kosice, Slovak Republic}

\affiliation [d] {Bogoliubov Laboratory of Theoretical Physics, Joint
Institute for Nuclear Research, 141980 Dubna, Moscow region, Russia}

\affiliation [e] {Division of Human Support System, Faculty of 
Symbiotic Systems Science, Fukushima University, Fukushima 960-1296, Japan}

\affiliation[f]{Dipartimento di Fisica, Universit´a di Napoli Federico II, I-80126 - 
Napoli, Italy}

\affiliation[g]{INFN Sez. di Napoli, Compl. Univ. di Monte S. Angelo,
Edificio G, I-80126 - Napoli, Italy}

\affiliation[h]{Gran Sasso Science Institute (INFN), Viale F. Crispi, 7, I-67100
L'Aquila, Italy}

\affiliation[j]{Instituto de F\'{\i}sica, Pontificia
Universidad de Cat\'olica de Valpara\'{\i}so, Casilla 4950,
Valpara\'{\i}so, Chile}

\affiliation[k]{CASPER, Physics Department, Baylor University, Waco, TX 76798-7310, USA}

\affiliation[l]{Physics Division,
National Technical University of Athens, 15780 Zografou Campus,
Athens, Greece}

\emailAdd{alireza.sepehri@uk.ac.ir}
\emailAdd{pincak@saske.sk}
\emailAdd{bamba@sss.fukushima-u.ac.jp}
\emailAdd{capozziello@na.infn.it}
\emailAdd{Emmanuel$_-$Saridakis@baylor.edu}

\abstract{We propose a model  describing the evolution of the free electron current 
density in 
graphene.
 Based on the concept of Mp-branes, we perform the analysis 
using the difference between curvatures of parallel and antiparallel spins. In such a 
framework an effective graviton emerges in the form of gauge field exchange between 
electrons. In a plain graphene system, the curvatures produced by both kinds of spins 
neutralize each other. However, in the presence of defects, the inequality between 
curvatures leads to the emergence of current density, modified gravity and conductivity. 
Depending on the type of the defects, the resulting current density can be negative or 
positive.}

\keywords{Graphene; Defects; M-theory; Modified gravity.}

\maketitle

\section{Introduction}

During the last years, graphene has been in the focus of frontier research, since its 
mass energy excitation for fermions is low, as it emerges from the symmetries of the 
honeycomb lattice \cite{s1,Haefner:2014sna,s2a,Ulybyshev:2015opa,s2b,s3,s4,s5}. This 
research has led to many important 
results. Amongst others, firstly, the surface free energy and the elastic constants 
(i.e. the Lame parameters such as Poisson ratio and Youngs modulus,) of graphene flakes,  
have been obtained on the level of the density functional theory. In particular, it has 
been shown that the Lame parameters in small graphene flakes may differ from the bulk 
values by thirty percent, for hydrogenated zig-zag edges which originate from the edge of 
the flake that compresses the interior \cite{s1}. Secondly, other investigations show 
that graphene with (effective) vacancy disorder is a physical representative of dirty 
d-wave superconductors \cite{Haefner:2014sna}. Thirdly, the density of states (DoS) of 
this system 
has been obtained numerically and within the self-consistent T-matrix approximation 
(SCTMA) in the presence of vacancies \cite{s2a,Ulybyshev:2015opa,s2b}. Fourthly, other 
works have considered 
quantum size effects in armchair graphene nanoribbons (AGNRs) with hydrogen termination 
via density functional theory (DFT) in Kohn-Sham formulation, and they have obtained the 
electronic structure of this system and predicted a threefold periodicity of the 
excitation gap with ribbon width \cite{s3}. Furthermore, in another scenario the local 
current density in pristine armchair graphene nanoribbons  with varying width has 
been calculated and it was found that the response of the current to functionalizing 
adsorbates is very sensitive to their placement: adsorbates located within the current 
filaments lead to strong backscattering, while adsorbates placed in other regions have 
almost no impact at all \cite{s4}. Moreover, in \cite{Kochetov:2010tm} it was shown that 
geometry has crucial effects on the electronic properties of the graphene. Also, the 
spin-dependent zero-bias conductance, in armchair graphene nanoribbons with hydrogen 
adsorbates, has been derived and the spin-orbit interaction has been considered. It has 
been observed that 
the spin-flip conductance can reach the same order of magnitude as the spin-conserving 
one, due to exchange-mediated spin scattering \cite{s5}. Additionally, in \cite{v5} 
the authors studied the electrically charged black holes in Palatini formalism of  $f(R)$ 
gravity, and they have argued that a form $f(R)= R\pm\alpha R^{2}/R_{P}$ (with $R_{P}$  
the Planck curvature) could induce different geometrical structures  in terms of inner 
black holes horizons. In a different scenario it was shown that Palatini  $f(R)$ gravity 
may lead to a geometry in which point-like space-time singularity in four dimensional 
classical models is replaced by a finite size wormhole structure, and this non-singular 
spacetime appears despite the existence of curvature divergences at the wormhole throat 
\cite{v6}. Finally, in \cite{v3}  the authors explained that microscopic wormholes which 
are created in modified gravity theories like $f(R)$ gravity have a role analogous to that 
of defects in crystals, which indicates the possible connection of solid-state and 
modified-gravity effects.

According to some considerations, massless two-dimensional Dirac fermions in a graphene 
behave in a way that can be described in terms of both quantum field theory and condensed 
matter physics. The formulated connection considers the role of gauge fields \cite{B1}, 
and motivated by this research we propose the model of the electronic transport in 
graphene in which Mp-branes 
\cite{Sepehri:2015ara,Sepehri:2016jfx,Sepehri:2014jla,Sepehri:2015eea, Sepehri:2016sjq, 
Sepehri:2015otm}, known from cosmology, appear. In 
the related superstring theory, initially there were only open scalar strings, which were 
later attached to zero dimensional objects, called M0-branes. This attachment is 
similar to the attachment of strings to a point-like one zero dimensional page. In fact, 
in string theory, each open string is attached from each end to one point or zero 
dimensional page which called M0. Moreover, at the beginning the shapes of all strings 
in respect to each other and to M0-branes are the same, they have no spin, and hence we 
cannot observe any difference between them. That is why we only have scalar strings at 
this stage. However, through interactions between strings and M0-branes, the shapes of 
strings in respect to each other are changed and they become different. Some of them gain 
properties of gauge fields with spin one,  some obtain properties of fields with 
spin 2, while some other achieve properties of other spins.  Nevertheless,  at 
the beginning, no gauge 
fields and fermions were present \cite{Sepehri:2016jfx}. Then, the M0-branes were joined 
and constructed a system of M1 and anti-M1-branes connected by a wormhole, a system
named BIon \cite{Sepehri:2014jla,Sepehri:2015eea,Sepehri:2016sjq,Sepehri:2015otm}. As a 
next step, the M0-branes glued to each other symmetrically, with the upper and lower parts 
of M1-brane being the same, and hence gauge fields were produced. Moreover, by linking 
M0-branes anti-symmetrically, fermions were created 
\cite{Sepehri:2016jfx,Sepehri:2016sjq}. Finally, these M1-BIons were 
connected and formed M3-BIons, which were a configuration of an M3, an anti-M3-brane 
connected by a wormhole \cite{Sepehri:2014jla,Sepehri:2015eea,Sepehri:2016sjq}. In this 
context/picture our Universe would be located on one of these M3-branes and its 
evolution is determined by interaction of branes in extra dimensions  
\cite{Sepehri:2015ara,Sepehri:2016jfx,Sepehri:2014jla,Sepehri:2015eea,Sepehri:2016sjq,
Sepehri:2015otm}.

Furthermore, there are works in the literature in which dark matter and cosmological 
acceleration are produced by braneworld scenarios. In particular, in \cite{Cembranos1} 
the authors showed  that in the context of brane-world models with low tension $\tau= 
f^{4}$, massive brane fluctuations are natural dark matter candidates, while  the 
present abundances for both hot(warm) and cold branons  were obtained in terms of the 
branon mass $M$ and the tension scale $f$. These results were in agreement with the recent 
experimental bounds on these parameters. Additionally, in a different braneworld 
model the main phenomenology associated with disformal scalars coupled to the Standard 
Model fields were considered, and  it was argued that these fields can serve as natural 
dark matter candidates since they are massive and weakly coupled \cite{Cembranos2}. 
Finally, in \cite{Cembranos3} the possible  detection of branon dark matter in 
experimental data was discussed.  For reviews on braneworld models which produce  
cosmological acceleration see \cite{braneworld1,braneworld2,braneworld3}. 

A similar model can be constructed  for  condensed matter systems. In this paper,  we 
propose an 
approach leading to  
  modified gravity \cite{Capozziello:2011et, Capozziello:2007ec} 
due to defects in graphene, 
and we calculate the current density of free electrons in terms of the curvature of the 
system. Each atom in graphene has three bound electrons, which are paired with the 
electrons 
of other atoms, and one free electron whose motion leads to the emergence of 
conductivity. 
Without free electrons,  there is a high symmetry in this system, and graphene behaves 
similarly to M0-brane, where paired electrons behave like scalar fields. However, by 
the motion of free electrons,  the system symmetry is broken and two  gauge 
fields are created. These gauge fields play the role of \textit{ gravitons} and produce 
the gravity 
 between anti-parallel spins and anti-gravity between parallel spins. For a graphene 
without defects, the curvature produced by non-identical spins is neutralized by the 
curvature of identical fermions, and the total curvature of the system becomes zero. This 
curvature ($R$) is related to the energy-momentum tensor 
$(T_{\mu}^{\nu}=diag[-p,-p,-p,\rho])$ by the relation ($T_{\mu\nu}=R_{\mu\nu}-\frac{1}{2}R 
g_{\mu\
nu}$)
and thus the imposed momentum to the electron is zero. Along this context, the 
meaning of anti-gravity becomes more transparent. In particular, when the 
curvature is zero the energy momentum tensor is zero ($R\approx T_{\mu}^{\nu}=0$) and 
consequently its components such as energy density and momentums are zero. Since force is 
related to momentum changes we deduce that the applied force to particles is zero. This 
implies that free electrons do not move in a certain path, and their current density 
becomes zero and thus 
conductivity disappears. On the other hand, the existence of suitable types of defects 
leads anti-parallel spins to come closer  mutually, and therefore their curvature 
increases. In this context,  modified gravity emerges and applies momentum to free 
electrons. Consequently, free electrons move in a special path, the current density 
increases and conductivity grows. Similarly, other types of defects make parallel spins 
to approach each other, thus increasing the curvature leading to the emergence of  
(modified) anti-gravity.  Contrary to the fact that gravity produces attractive 
force, anti-gravity creates a repulsive force between bodies, objects and particles. In 
anti-gravity case, momentum is applied to free electrons in opposite 
directions, and hence the sign of current density reverses. When the curvature  is 
negative, it's related effective energy-momentum tensor  is 
negative and consequently, the applied force becomes negative. Consequently, particles 
move mutually away.

The outline of the paper is the  following. In Section \ref{o1} we obtain the current 
density in terms of inequality between curvatures of parallel and anti-parallel spins.   
In Section \ref{o2} we show that this density is zero for standard graphene, positive for 
pentagonal defects and negative for heptagonal defects. In Section \ref{o3} we obtain the 
current density in graphene wormholes. Finally, Section  \ref{sum} is 
devoted to discussion and conclusions.

\section{The current density  in graphene}
\label{o1}

Let us  introduce the  concepts of Mp-branes in order to extract  the 
current density of free electrons in terms of inequality between curvatures of parallel 
spins and anti-parallel spins. In order to achieve this result, we apply the 
method used in \cite{Sepehri:2015ara,Sepehri:2016jfx,Sepehri:2016sjq} for calculating the 
energy in terms of curvature, and we
write the explicit form of curvatures in terms of couplings of parallel and anti-parallel 
spins. 

Let us begin by considering  the scalar fields $X^{M}=X^{M}_{\alpha}T^{\alpha}$, where 
the  generators $T^{\alpha}$ of the  Lie 3-algebra    
satisfy the 3-dimensional Nambu-Poisson brackets 
\cite{Bagger:2007jr,Myers:1999ps,Gustavsson:2007vu,Constable:2001ag,Sepehri:2016nuv,
Ho:2008nn, Mukhi:2008ux} 
 \begin{eqnarray}
  &&[T^{\alpha}, T^{\beta}, T^{\gamma}]=
  f^{\alpha \beta \gamma}_{\eta}T^{\eta} \nonumber \\&& [X^{M},X^{N},X^{L}]=
  [X^{M}_{\alpha}T^{\alpha},X^{N}_{\beta}T^{\beta},X^{L}_{\gamma}T^{\gamma}],
 \label{w3}
 \end{eqnarray}
with $f^{\alpha \beta \gamma}_{\eta}$ the structure constants. Let us also consider   a
2-form gauge   field $A_{ab}$. In this case, the full action in M-theory reads
\cite{Sepehri:2015ara,Sepehri:2016jfx,Sepehri:2016sjq}:
\begin{eqnarray}
  &&S_{co-Graphene} = \int  d^{3}x \sum_{n=1}^{U}\beta_{n}\Big(  
\delta^{a_{1},a_{2}...a_{n}}_{b_{1}b_{2}....b_{n}}L^{b_{1}}_{a_{1}}...L^{b_{n}}_{a_{n}}
\Big)^{1/2},
\label{action1}
  \end{eqnarray}
  where 
  \begin{eqnarray}
  (L)^{a}_{b}=\delta_{a}^{b}  STr \left\{-det\left(P_{abc}[ E_{mnl}
  +E_{mij}(Q^{-1}-\delta)^{ijk}E_{kln}]+ \lambda
  F_{abc}\right)det(Q^{i}_{j,k})\right\}\label{w1},
  \end{eqnarray}
with 
 \begin{eqnarray}
  E_{mnl}^{\alpha,\beta,\gamma} &=& G_{mnl}^{\alpha,\beta,\gamma} + 
B_{mnl}^{\alpha,\beta,\gamma}, \nonumber\\
  Q^{i}_{j,k} &=& \delta^{i}_{j,k} + 
i\lambda[X^{j}_{\alpha}T^{\alpha},X^{k}_{\beta}T^{\beta},X^{k'}
_{\gamma}T^{\gamma}]
  E_{k'jl}^{\alpha,\beta,\gamma},\nonumber\\F_{abc}&=&\partial_{a} A_{bc}-\partial_{b} 
A_{ca}+\partial_{c}
 A_{ab}. \label{w2}
 \end{eqnarray}
  In the above expressions 
 $G_{mnl}=g_{mn}\delta^{n'}_{n,l}+\partial_{m}X^{i}\partial_{n'}X^{i}\sum_{j}
 (X^{j})^{2}\delta^{n'}_{n,l} + \frac{1}{2}\langle 
\partial^{b}\partial^{a}X^{i},\partial_{b}\partial_{a}X^{i}\rangle $, 
  $P_{abc}$ is the pull-back of scalars, and 
$STr$ stands for the symmetric trace of products. In graphene, the scalar fields denote 
the pairs which are produced by pairing anti-parallel electrons. Similarly,
$\lambda=2\pi l_{s}^{2}$ where, in M-theory,  $l_{s}$ is the string length, which, 
in graphene, is just the separation distance between atoms. Additionally, $\beta_{n}$ 
are constants, related to different atoms in graphene. Concerning the index notation, 
note that, in graphene,  $a,b$ denote indices of pairs on each atom, while 
$i,j$ refer to indices of free pairs corresponding to free electrons. Finally,  $U$ 
denotes the number of atoms and $p$ is the 
number of pairs in each atom.

Applying action (\ref{action1}) to graphene,  we assume that the scalar fields $ X^{i} $ 
play the role of pairs of anti-parallel electrons and the 2-form gauge tensor  fields 
$A_{ab}$
play the role of gravitons which are exchanged between electrons. This assumption  is
 justified since, in graphene, the three electrons of each atom are paired with three 
electrons of another atom by exchanging 2-form gauge fields  forming spinless pairs which 
 can be treated as scalars. Moreover, free electrons are represented by $\psi$, while 
electrons in each pair are denoted by $\Psi$. Hence, with these positions,  we 
can define \cite{Sepehri:2016sjq}:
 \begin{eqnarray}
 && A_{ab}\rightarrow \psi^{U}_{a}\psi^{L}_{b}-\psi^{L}_{a}\psi^{U}_{b}\nonumber \\
  && X\rightarrow 
\psi^{U}_{a}A^{ab}\psi^{L}_{b}-\psi^{L}_{a}A^{ab}\psi^{U}_{b}+\Psi^{U}_{a}A^{ab}\Psi^{L}_{
b}-\Psi^{L}_{a}A^{ab}\Psi^{U}_{b}+\Psi^{U}_{a}A^{ab}\psi^{L}_{b}-\psi^{L}_{a}A^{ 
ab}\Psi^{U}
_{b}\nonumber \\
 &&\partial_{a} =\partial_{a}^{U}+\partial_{a}^{L} \nonumber \\
  &&\partial_{a}^{U}\psi^{U}_{a}=1,\;\partial_{a}^{L}\psi^{L}_{a}=1.
  \label{w4}
 \end{eqnarray}
Using these definitions we can make the splitting  \cite{Sepehri:2016sjq}:
 \begin{equation}
 \langle
 F^{abc},F_{abc}\rangle_{tot}\equiv\langle
 F^{abc},F_{abc}\rangle_{Free-Free}+\langle
 F^{abc},F_{abc}\rangle_{Free-Bound}+\langle
 F^{abc},F_{abc}\rangle_{Bound-Bound},
 \label{w5}
 \end{equation}
where the subscript  ``Free-Free'' indicates the mutual interaction of two free 
electrons, 
``Free-Bound''  denotes the mutual interaction of free and bound electrons, and 
``Bound-Bound'' denotes the mutual interaction of two bound electrons. Thus, 
using the  definitions (\ref{w4}), we can calculate the different terms of (\ref{w1}) in 
terms of couplings of parallel and anti-parallel spins \cite{Sepehri:2016sjq}. The 
corresponding expressions are shown in Appendix \ref{Appterms}.
 
In summary,  from the above analysis, by breaking pairs into electrons,   
standard Dirac equations  are obtained, and hence,  the relation between 
gauge fields and their sources, i.e. fermions, becomes clear. These results are very 
similar to those in M-theory \cite{Sepehri:2015ara,Sepehri:2016jfx,Sepehri:2016sjq}. 
Furthermore, one can use these 
expressions to show that, by joining electrons and forming a pair, 2-form gauge fields 
are 
created: these gauge fields  play the role of  graviton tensor modes between two electrons 
($\Psi^{\dag 
a,U}\langle F_{abc},F^{i'bc}\rangle\psi_{i'}^{L}$). Specifically,  one can  obtain the 
couplings 
between electrons in terms of \textit{curvature} of 
graphene. As it was shown in 
\cite{Sepehri:2015ara,Sepehri:2016jfx,Sepehri:2016sjq,Mannheim:2014gba,Ni:2014qfa},  the 
metric can be 
antisymmetric, since, for instance,   M1-branes are linked 
to anti-M1-branes and form a 
new system \cite{Sepehri:2015ara,Sepehri:2016jfx,Sepehri:2016sjq}. Thus, for this system 
the metric can be constructed from 
metrics 
of two M1's as:
\begin{equation}
 \text{Metric of 
system}\equiv(\text{Metric M1})_{1}\otimes (\text{Metric M1})_{2}-(\text{Metric 
M1})_{2}\otimes (\text{Metric 
M1})_{1},
 \end{equation}
and therefore it can be antisymmetric. On the other hand, the graviton tensor mode 
has a direct relation with the metric and it can be anti-symmetric. 
 
 The same conditions can be realized in the graphene.  In particular,  the 
metric in a graphene  can 
be constructed from the metric of pairs, such as for two electrons of a pair: 
\begin{eqnarray}
&&\!\!\!\!\!\!\!\!\!\!\!\!\!\!\!\!\!\!\!\!\!\!\!\!\!\!\!\!\!\!\!\!
\text{Metric of pair}\equiv(\text{
Metric electron})_{1}\otimes (\text{Metric electron})_{2}\nonumber\\
&&\ \ \ \ \ \  
-(\text{Metric 
electron})_{2}\otimes (\text{Metric electron})_{1}.
\end{eqnarray}
Since this metric can be antisymmetric,  the tensor mode of the graviton 
may be antisymmetric. Hence, the 2-form gauge fields have a direct relation with the
graviton, and also with the metric of pairs in the graphene. 

Using the aforementioned method in a graphene structure, we can obtain the relation 
between 
fermions and curvatures \cite{Sepehri:2015ara,Sepehri:2016jfx,Sepehri:2016sjq}:
\begin{eqnarray}
 &&A^{ab}=g^{ab}=h^{ab}=h_{1}^{ab'}\otimes h_{2}^{b'b}-h_{2}^{bb'}\otimes h_{1}^{ab'} 
 \nonumber\\&& 
 F_{abc}=\partial_{a} A_{bc}-\partial_{b}
 A_{ca}+\partial_{c} 
A_{ab}=2(\partial_{\mu}g_{\nu\lambda}+\partial_{\nu}g_{\mu\lambda}-\partial_{\lambda}g_{
\mu\nu})=
 2\Gamma_{\mu\nu\lambda}  
  \end{eqnarray}
  \begin{eqnarray}
 &&\!\!\!\!\!\!\!\!\!\!\!\!\!\!\!\!\!\!\!\!\!\!\!\!
 \langle
 F^{\rho}\smallskip_{\sigma\lambda},F^{\lambda}\smallskip_{\mu\nu}\rangle=
\langle[X^{\rho},X_{\sigma},X_{\lambda}],[X^{\lambda},X_{\mu},X_{\nu}]
\rangle
\nonumber\\
&&\ \ \ \ \ \  
=
[X_{\nu},[X^{\rho},X_{\sigma},X_{\mu}]]-[X_{\mu},[X^{\rho},X_{\sigma},X_{\nu}]]
\nonumber\\
&&\ \ \ \ \ \  \  \  \  \, +[X^{\rho}
,
X_{\lambda},X_{\nu}]
 [X^{\lambda},X_{\sigma},X_{\mu}]
-[X^{\rho},X_{\lambda},X_{\mu}][X^{\lambda},X_{\sigma},X_{\nu}]
\nonumber\\
 &&\ \ \ \ \ \   =
\partial_{\nu
}
\Gamma^{\rho}_{\sigma\mu}-\partial_{\mu}\Gamma^{\rho}_{\sigma\nu}+\Gamma^{\rho}_{
\lambda\nu}
 \Gamma^{\lambda}_{\sigma\mu}-\Gamma^{\rho}_{\lambda\mu}\Gamma^{\lambda}_{\sigma\nu}
 =R^{\rho}_{\sigma\mu\nu},
 \end{eqnarray}
where
$h^{ab}$ is the metric which is seen by the electron, $X_{\sigma}$ is the scalar which is 
created by pairing two anti-parallel electrons, and $R^{\rho}_{\sigma\mu\nu}$ is the 
Riemann tensor. Additionally, we obtain
  \begin{equation}
  \langle
     F_{abc},F_{a'}^{bc}\rangle=R_{aa'}^{anti-parallel}-R_{aa'}^{parallel},
      \end{equation}
      and
       \begin{eqnarray}
  &&
  R_{MN}=R_{
aa'}+R_{ia'}+R_{ij'}=R_{Free-Free}^{anti-parallel}+R_{Free-Bound}^{anti-parallel}+R_{
Bound-Bound}^{
anti-parallel}\nonumber\\
&&
\ \ \ \ \ \ \ \ \ \ \ 
-R_{Free-Free}^{parallel}-R_{Free-Bound}^{parallel}-R_{
Bound-Bound}^{
parallel}.
\label{w7}
 \end{eqnarray}
 Finally, the term $\langle 
\partial^{b}\partial^{a}X^{i},\partial_{b}\partial_{a}X^{i}\rangle$ is given in Appendix 
\ref{Bigfermcurvterm}. In the above expressions,  $R_{Bound-Bound}^{anti-parallel}$ is 
the 
curvature produced by the interaction of two bound anti-parallel electrons, 
$R_{Bound-Bound}^{parallel}$ is the curvature created by interaction of two bound 
parallel 
electrons, $R_{Free-Free}^{anti-parallel}$ is the curvature produced by the interaction 
of 
two free anti-parallel electrons, $R_{Free-Free}^{parallel}$ is the curvature created by 
interaction of two free parallel electrons, $R_{Free-Bound}^{anti-parallel}$ is the 
curvature produced by the interaction of  free and bound anti-parallel electrons and 
$R_{Free-Free} ^{parallel}$ is the curvature created by interaction of  free and bound 
parallel electrons. 
                          
In summary, from the above expressions,  we deduce  that there are two types of effective 
gravity: one 
related to parallel spins and one related to anti-parallel spins. Moreover, we have three 
types of curvatures: A first type of curvature is created by mutual coupling 
of free electrons. A second type of curvature  is produced by coupling of free 
electrons to bound electrons. Lastly, a third type of curvature is produced by mutual
coupling of bound electrons. Note that the curvature between anti-symmetric fermions  
is positive, while the curvature between parallel spins is negative. 
In fact, the gravity between anti-parallel spins, namely
$\Psi^{\dag a,U}R_{aa'}^{anti-parallel} \Psi^{a',L}$, creates an attracting force, 
while the anti-gravity between parallel 
spins, i.e. -$\Psi^{\dag a,U}R_{aa'}^{parallel} \Psi^{a',U}$, produces a repelling 
force. This is explained since curvature has a direct relation with energy-momentum 
tensor, namely  $\Psi^{\dag a,U}R_{aa'}^{parallel} \Psi^{a',U}\approx \Psi^{\dag 
a,U}T_{aa'}^{parallel} \Psi^{a',U}$. Since one of the components of this tensor is the 
momentum, i.e. $T_{aa'}\propto P$, by changing the curvature of the system, the  momentum 
changes and attractive or repulsive force  emerge. Schematically we have: 
    \begin{eqnarray}
  && \! -\Psi^{\dag a,U}R_{aa'}^{parallel} \Psi^{a',U}\approx -\Psi^{\dag 
a,U}T_{aa'}^{parallel} \Psi^{
a',U} \propto -\Psi^{\dag a,U}\vec{P} \Psi^{a',U}
\nonumber \\&&  
\!\!\!\!\!\!\!\!
\Rightarrow
-\frac{d}{dt}(\Psi^{\dag a,U}R_{aa'}^{parallel} \Psi^{a',U})\approx 
-\frac{d}{dt}(\Psi^{\dag a,U}\overrightarrow{P} \Psi^{a',U})\approx -\Psi^{\dag a,U}F 
\Psi^{a',U}+\cdots,
  \end{eqnarray}
  and
    \begin{eqnarray}
  && \! \Psi^{\dag a,U}R_{aa'}^{anti-
parallel} \Psi^{a',L}\approx \Psi^{\dag a,U}T_{aa'}^{anti-parallel} \Psi^{a',L} \propto 
\Psi^{\dag 
a,U}\overrightarrow{P} \Psi^{a',L}
\nonumber \\&&  
\!\!\!\!\!\!\!\!
\Rightarrow
\frac{d}{dt}(\Psi^{\dag 
a,U}R_{aa'}^{
anti-parallel} \Psi^{a',L})\approx \frac{d}{dt}(\Psi^{\dag a,U}\vec{P}  
\Psi^{a',L})\approx \Psi^{\dag a,U}F \Psi^{a',L}+\cdots . 
\label{w9}
  \end{eqnarray}
  From the above analysis,  we deduce that positive curvature between anti-parallel spins 
leads to an attractive force between them, while negative curvature between parallel 
spins causes a mutual repulsive force. These forces produce the current density in the 
system, which we are interested in calculating. 

Let us  start by considering the 
interaction between electrons of one atom with those of neighbor atoms, and then we 
generalize it to the total system. 
Substituting expressions (\ref{w7}) and (\ref{w8}) in Eq. (\ref{w1}), for one atom 
of 
graphene (n=1), after some algebra,   we extract the energy as:
\begin{eqnarray}
 && \!\!\!\!\!\!\!\!\!\!\!\!\!
 E_{system}=\int d^{4}x \rho
 \nonumber
 \\
 &&\ \ \ \
 =     \int d^{4}x   \sqrt{-g}\,
      \Bigg\{
     -(1-m_{g}^{2})
                \Big[( R_{Free-Free}^{parallel})^{2}+( 
R_{Free-Free}^{anti-parallel})^{2}+( 
R_{Free-
Bound}^{parallel})^{2} 
\nonumber 
\\
&& \ \ \ \ \ \ \ \ \ \ \ \ \ \ \ \ \  \  \ \
+
( R_{Free-Bound}^{anti-parallel})^{2}+( 
R_{Bound-Bound}^{
parallel})^{2}+( R_{Bound-Bound}^{anti-parallel})^{2}
\nonumber
\\
&& \ \ \ \ \ \ \ \ \ \ \ \ \ \ \ \ \  \  \ \  + 
(R_{Free-Free}^{parallel} R_{
Free-Free}^{anti-parallel})\,\partial^{2}(R_{Free-Free}^{parallel}+ 
R_{Free-Free}^{anti-parallel} )
\nonumber \\
&& \ \ \ \ \ \ \ \ \ \ \ \ \ \ \ \ \  \  \ \ 
+(R_{Free-Bound}^{parallel} 
R_{Free-Bound}^{anti-parallel})\,\partial^{2}(R_{Free-Bound}^{
parallel}+ R_{Free-Bound}^{anti-parallel} )
\nonumber \\
&& \ \ \ \ \ \ \ \ \ \ \ \ \ \ \ \ \  \  \ \ 
+
(R_{Bound-Bound}^{parallel} 
R_{Bound-
Bound}^{anti-parallel})\,\partial^{2}(R_{Bound-Bound}^{parallel}+ 
R_{Bound-Bound}^{anti-parallel} ) \Big]
\nonumber\\
&& \ \ \ \ \ \ \ \ \ \ \ \ \ \ \ \ \ 
+   m_{g}^{2}\lambda^{2}\delta_{\rho_{1}\sigma_{1}}^{\mu_{1}\nu_{1}}              
\Big[
R^{anti-parallel,\rho_{1}\sigma_{1}}_{Free-Free,\mu_{1}\nu_{1}}+R^{anti-parallel,
\rho_{1}\sigma_{1}}_{Bound-Bound,\mu_{1}\nu_{1}}+R^{anti-parallel,\rho_{1}\sigma_{1}}_{
Free-Bound,\mu_{1}\nu_{1}}
\nonumber\\
&& \ \ \ \ \ \ \ \ \ \ \ \ \ \ \ \ \  \  \ \ 
-
R^{parallel,\rho_{1}\sigma_{1}}_{Free-Free,\mu_{
1}\nu_{1}} +R^{
parallel,\rho_{1}\sigma_{1}}_{Bound-Bound,\mu_{1}\nu_{1}}+R^{parallel,\rho_{1}\sigma_{1}}_
{Free-
Bound,\mu_{1}\nu_{1}}
\Big]  
\Bigg\},
\label{w10}
    \end{eqnarray}
 where
$m_{g}^{2}=(\lambda)^{2}\det([X^{j}_{\alpha}T^{\alpha},X^{k}_{\beta}T^{\beta},X^{k'}_{
\gamma}
T^{\gamma}])$ is the  graviton mass square. 
    Note that,  for the sake of  simplicity,   we have chosen 
        \begin{equation}
     \Psi^{\dag a, 
L}\psi^{U}_{a}=\Psi^{\dag a, U}\psi^{L}_{a}=\Psi^{\dag a, L}\Psi^{U}_{a}=\Psi^{\dag a, 
U}\Psi^{L}_{a}=\psi^{\dag a, 
U}\psi^{L}_{a}=\psi^{\dag a, L}\psi^{U}_{a}=l_{1},
\label{choicepsi1}
    \end{equation}
 and 
     \begin{equation}
  \Psi^{\dag a, 
U}\psi^{U}_{a}=\Psi^{\dag a, 
L}\psi^{L}_{a}=\Psi^{\dag a, U}\Psi^{U}_{a}=\Psi^{\dag a, L}\Psi^{L}_{a}=\psi^{\dag a, 
U}\psi^{U}_{
a}=\psi^{\dag a, L}\psi^{L}_{a}=l_{2},
\label{choicepsi2}
    \end{equation}
   where $l_{1}$ is the coupling between two 
anti-parallel spins and $l_{2}$ the coupling between parallel spins.
Under these considerations,  we can  calculate  the curvatures 
as
 \begin{eqnarray}
                   && 
R_{Free-Free}^{parallel}=R_{Free-Bound}^{parallel}=R_{Bound-Bound}^{parallel}\approx 
l_{2}-l'_{2}\nonumber \\&&R_{Free-Free}^{anti-parallel}=R_{Free-Bound}^{anti-parallel}=R_{
Bound-Bound}^{anti-parallel}\approx l_{1}+l'_{1}.
\label{curv11}
     \end{eqnarray}
Thus, using these useful relations, together with (\ref{w10})  and the chain rule, we 
obtain  
      \begin{eqnarray}
          &&\frac{\partial 
E_{system}}{\partial t}
=\frac{\partial E_{system}}{\partial l_{1}}\frac{\partial l_{1}}{\partial 
t}+\frac{\partial E_{
system}}{\partial l_{2}}\frac{\partial l_{2}}{\partial t}\,.
     \end{eqnarray}
 Finally, we get  the 
current density as:
 \begin{eqnarray}
 && \!\!\!\!\!\!\!\!\!\!\!\!
 I=\int 
d^{4}x J \approx \frac{\partial E_{system}}{\partial t}
\nonumber 
\\&&\!\!\!\!\!\!\!\!
\approx 
-   \int d^{4}x   \sqrt{-g}\,
      \Bigg\{    
      -(1-m_{g}^{2})
         \Big[( R_{Free-Free}^{parallel})^{2}-( 
R_{Free-Free}^{anti-parallel})^{2}+( 
R_{Free-
Bound}^{parallel})^{2}
\nonumber \\
&& \ \ \ \ \ \ \ \ \ \ \ \ \ \  \ \  \ \
-
( R_{Free-Bound}^{anti-parallel})^{2}+( 
R_{Bound-Bound}^{
parallel})^{2}-( R_{Bound-Bound}^{anti-parallel})^{2}
\nonumber \\
&& \ \ \ \ \ \ \ \ \ \ \ \ \ \  \ \  \ \ 
+ 
(R_{Free-Free}^{parallel} R_{
Free-Free}^{anti-parallel})\,\partial^{2}(R_{Free-Free}^{parallel}- 
R_{Free-Free}^{anti-parallel} )
\nonumber 
\\
&& \ \ \ \ \ \ \ \ \ \ \ \ \ \  \ \  \ \ 
+
(R_{Free-Bound}^{parallel} 
R_{Free-Bound}^{anti-parallel})\,\partial^{2}(R_{Free-Bound}^{
parallel}- R_{Free-Bound}^{anti-parallel} )
\nonumber
\\
&& \ \ \ \ \ \ \ \ \ \ \ \ \ \  \ \  \ \  
+
(R_{Bound-Bound}^{parallel} 
R_{Bound-
Bound}^{anti-parallel})\,\partial^{2}(R_{Bound-Bound}^{parallel}- 
R_{Bound-Bound}^{anti-parallel} )
 \Big]
\nonumber\\
&& \ \ \ \ \ \ \ \ \ \ \ \ \ \  
+
m_{g}^{2}\lambda^{2}\delta_{\rho_{1}\sigma_{1}}^{\mu_{1}\nu_{1}}        
   \Big[ R^{anti-parallel,\rho_{1}\sigma_{1}}_{Free-Free,\mu_{1}\nu_{1}}+R^{anti-
parallel,\rho_{1}\sigma_{1}}_{Bound-Bound,\mu_{1}\nu_{1}}+R^{anti-parallel,\rho_{1}\sigma_
{1}}_{
Free-Bound,\mu_{1}\nu_{1}}
\nonumber\\
&& \ \ \ \ \ \ \ \ \ \ \ \ \ \  \ \  \ \ 
+
R^{parallel,\rho_{1}\sigma_{1}}_{Free-Free,\mu_{
1}\nu_{1}}
+R^{parallel,\rho_{1}\sigma_{1}}_{Bound-Bound,\mu_{1}\nu_{1}}+R^{parallel,\rho_{1}\sigma_{
1}}_{Free-
Bound,\mu_{1}\nu_{1}}\Big]
\Bigg\}.
\label{w11}
                  \end{eqnarray}
 From this expression, it is easy to see that the current density in graphene depends on 
the 
curvature produced by parallel and anti-parallel spins. For a symmetric graphene, 
curvatures of parallel spins are canceled by curvatures of anti-parallel spins, and 
therefore the current density of the system decreases. Hence, by increasing the 
symmetry conductivity decreases.
        
Finally, in order to calculate the total current density in graphene, we should sum over 
currents in each atom, namely    
    \begin{eqnarray}
 \vec{J}
_{system}= 
\Sigma_{n=1}^{U}\delta^{a_{1},a_{2}...a_{n}}_{b_{1}b_{2}....b_{n}}\vec{J}^{b_{1}}_{a_{1}}
...\vec{J}^{b_{n}}_{a_{n}},
\label{w12}
    \end{eqnarray}
   where $\vec{J}^{b_{n}}_{a_{n}}=\vec{J}\delta^{b_{n}}_{a_{n}}$. 
 This expression indicates that the total current density in graphene depends on the 
curvatures of parallel spins and anti-parallel spins in each atom. If the current density 
in one atom is zero, electrons stop at that point and the total current density of the 
system becomes zero. This fact  leads to the disappearance of conductivity. Thus, by 
breaking the 
symmetry in graphene, curvatures are created. This phenomenon  leads to the production of 
current 
density and hence of conductivity in the system.

\section{The current density in graphene in presence of various defects}
\label{o2}

In this section,  we investigate  the  model in the specific case of standard  graphene, 
and, 
additionally,  considering   heptagonal and pentagonal defects. In particular, using the 
definitions for couplings of parallel spins ($l_{2}$) and for couplings of 
anti-parallel spins ($l_{1}$), and  substituting expressions (\ref{w7}) and (\ref{w8})
into expression (\ref{w1}), we obtain the following action for one atom in the graphene:
\begin{eqnarray}
   && 
   \!\!\!\!\!\!\!\!\!\!\!\!\!\!\!\!\!
   S_{co-atom}\approx V\int  d\cos\theta\,\Sigma_{n=1}^{p} 
\Big\{6 m_{g}^{2}\lambda^{2}\left[l_{1}-
l_{2}-l'_{1}+l'_{2}+(l'_{1})^{2}-(l'_{2})^{2}\right]
\nonumber
\\
&& \ \ \ \ \ \ \ \ \  \ \ \ \ \ \ \ \  \ \ \ \ \ \ \ \ \ 
-3(1-m_{g}^{2})
\left[2l_{1}^
{2}+2l_{2}
^{2}+2(l'_{1})^{2}+2(l'_{2})^{2}+l_{1}^{2}l_{2}^{2}(l_{1}^{2}+l_{2}^{2})''\right] 
\Big\}^{\frac{1}{2}},
\label{w13}
       \end{eqnarray}
 with $V$ is the atom volume, and where  we have assumed that the 
couplings  depend only on $\theta$, i.e.  the angle between two electrons in one graphene 
atom with respect to the centre of the atom, with $'$ denoting derivative with respect to 
$\cos \theta$.

The variation of the action  (\ref{w13}) provides the equations of motion as:
          \begin{eqnarray}          
&&
\!\!\!\!\!\!\!
\Big\{
m_{g}^{2} l'_{1}   \left[  \lambda^{2}-(1+2l_{1}^{2}l_{2}^{2})
\right]
\left\{
6 m_{g}^{2}\lambda^{2}[l_{1}-l_{2}-l'_{1}+l'_{2}+(l'_{1})^{2}-(l'_{2})^{2}]
\right.
\nonumber\\
&& \ \ \ \ \ \ \ \
\left.-3(
1-m_{g}^{2})
\left[2l_{1}^{2}+2l_{2}^{2}+2(l'_{1})^{2}+2(l'_{2})^{2}+l_{1}^{2}l_{2}^{2}(l_{1}^{
2}+l_{2}^{
2})''\right]\right\}^{-\frac{1}{2}}\Big\}'
=
\nonumber
\\
&& \!\!\!\!\!\!\!
\left\{(1-m_{g}^{2})l_{1}
\left[1
+3 l_{2}^{2}(l_{1
}^{2}+l_{
2}^{2})''\right]+m_{g}^{2}\lambda^{2}
\right\}
\left\{ 6  m_{g}^{2}\lambda^{2}[l_{1}-l_{2}-l'_{1}+l'_{2}
+(l'_{1}
)^{2}-(l'_{2})^{2}]
\right.
\nonumber\\
&& \ \ \ \ \ \ \ \
\left.
-3(1-m_{g}^{2})[2l_{1}^{2}+2l_{2}^{2}+2(l'_{1})^{2}+2(l'_{
2})^{2}
+l_{1}^{2}l_{2}^{2}(l_{1}^{2}+l_{2}^{2})'']\right\}^{-\frac{1}{2}},
\label{w14}
          \end{eqnarray}    
          \begin{eqnarray}          
&&
\!\!\!\!\!\!\!
\Big\{l'_{2}
\left[(1-m_{g}^{2})
(1-2l_{1}^{2}l_{2}^{2})
-m_{g}^{2}\lambda^{2}
\right]
\left\{
6m_{g}^{2}\lambda^{2}[l_{1}-l_{2}-l'_{1}+l'_{2}+(l'_{1})^{2}-(l'_{2})^{2}
]\right.
\nonumber
\\
&&
 \ \ \ \ \ \ \ \
\left.
-3(
1-m_{g}^{2})[2l_{1}^{2}+2l_{2}^{2}+2(l'_{1})^{2}+2(l'_{2})^{2}+l_{1}^{2}l_{2}^{2}(l_{1}^{
2}+l_{2}^{
2})'']\right\}^{-\frac{1}{2}}\Big\}'
=
\nonumber\\
&& \!\!\!\!\!\!\!
\left\{(1-m_{g}^{2})l_{2}
[1-3l_{1}^{2}(l_{
1
}^{2}+l_{
2}^{2})'']
-m_{g}^{2}\lambda^{2}
\right\}
\left\{
6 m_{g}^{2}\lambda^{2}[l_{1}-l_{2}-l'_{1}+l'_{2}
+(l'_{1}
)^{2}-(l'_{2})^{2}]
\right.
\nonumber
\\
&& \ \ \ \ \ \ \ \
\left.
-3(1-m_{g}^{2})[2l_{1}^{2}+2l_{2}^{2}+2(l'_{1})^{2}+2(l'_{
2})^{2}
+l_{1}^{2}l_{2}^{2}(l_{1}^{2}+l_{2}^{2})'']
\right\}^{-\frac{1}{2}}.
\label{w15}
      \end{eqnarray}
One can easily find the approximate solution of these equations as:   
    \begin{eqnarray}
    && l_{1}\approx \cos(\theta_{1}) \nonumber\\
    &&
    l_{2}\approx \cos(\theta_{2})= 
(1-m_{g}^{2})\cos(\theta_{
1})-m_{g}^{2}\lambda^{2}\sin(\theta_{1}),
\label{w16a}
      \end{eqnarray}
    and therefore expressions (\ref{choicepsi1}),(\ref{choicepsi2}) lead to
     \begin{eqnarray}
\Psi=\psi\approx\sqrt{\cos(\theta)}.
\label{w16}
  \end{eqnarray}
     Solutions (\ref{w16a}),(\ref{w16}) allow us to calculate the curvatures through 
(\ref{curv11}),  and finally  the  current density through (\ref{w11}).

\subsection{Graphene without defects (hexagonal)}

Let us first study the  simple case of  hexagonal graphene. It has been 
shown that, in the graphene,  the difference between angles of couplings of parallel 
spins 
and anti-parallel spins is around $\frac{\pi}{3}$ \cite{B14a,Gonzalez:2009je,B14b}, a 
result that is in agreement with the 
symmetries of the graphene. In each hexagonal graphene molecule,  the electrons of each 
atom should be anti-parallel with respect to the electrons of neighbor atoms, and 
therefore the angle between them with respect to  the molecule center is $\frac{\pi}{3}$. 
On the other hand, the angle between parallel spins is $\frac{2\pi}{3}$. 
Thus, the difference between angles of couplings of parallel spins and anti-parallel 
spins 
is $\frac{\pi}{3}$.
  Hence,  choosing the parameter values  $m_{g}^{2}=\frac{1}{2}$ and 
$m_{g}^{2}\lambda^{2}=\sqrt{3}$, we get
    \begin{eqnarray}
    && l_{1}\approx \cos(\theta_{1}) \nonumber\\
    &&
    l_{2}\approx 
\cos\left(\theta_{1}+\frac{\pi}{3}\right),
\label{w17}
      \end{eqnarray}
 which is exactly this realization.
      
\begin{figure}[ht]
\centering
\includegraphics[width=0.35\linewidth]{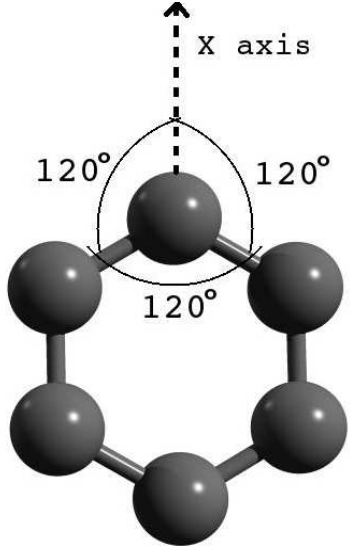}
\caption{{\it{The angle between electron pairs in the graphene without defects 
(hexagonal).}}}
\label{hexagonal}
\end{figure}

In order to calculate the action of conductivity in the
graphene, we should first obtain the  values of the curvature tensor at 
the places of the atoms. Each carbon has three bound electrons and one free electron, 
where the bound states are located along three axes that form an angle of $2\pi/3$.  The 
angle between two electron spins has a direct relation to the radius of the hexagonal 
molecule ($R$) and the separation distance between two electrons ($L$) 
($\theta=\frac{L}{ R}$ in radians). Thus, if we calculate the curvature in 
terms of the angle, we can obtain the explicit form of it in terms of the other 
parameters of the graphene. We assume that one axis lies along $X$ (see Fig. 
\ref{hexagonal}), and additionally we consider that the couplings for free and bound 
electrons are the same. Thus, we obtain the couplings of parallel and anti-parallel spins 
in terms of their angle with respect to the $X$-axis, namely  
      \begin{eqnarray}
      &&\!\!\!\!\!\!\!\!
      l_{1}^{1-1}\approx \cos(0)=1, \quad l_{1}^{1-2}\approx \cos(2\pi/3)=-\frac{1}{2},
\quad l_{1}^{
1-3}\approx \cos(4\pi/3)=-\frac{1}{2},
\nonumber\\
   &&\!\!\!\!\!\!\!\!
(l'_{1})^{1-1}\approx \sin(0)=0, \quad 
(l'_{1})^{1-2}
\approx \sin(2\pi/3)=\frac{\sqrt{3}}{2}, \quad (l'_{1})^{1-3}\approx 
\sin(4\pi/3)=-\frac{\sqrt{3}}{2},
\label{w18a}
        \end{eqnarray}
        and
  \begin{eqnarray}
      &&\!\!\!\!\!\!\!\!
      l_{2}^{1-1}\approx 
\cos(\pi/3)=\frac{1}{2},  \quad l_{2}^{1-2}\approx \cos(
\pi)=-1,
\quad l_{2}^{1-3}\approx \cos(5\pi/3)=\frac{1}{2},
\nonumber\\
   &&\!\!\!\!\!\!\!\!
(l'_{2})^{1-1}\approx 
\sin(\pi/3)=\frac{\sqrt{3}}{2},
\quad (l'_{2})^{1-2}\approx \sin(\pi)=0,
\quad 
(l'_{2})^{1-3}\approx \sin(5\pi/3)=-\frac{\sqrt{3}}{2}.
\label{w18}
\end{eqnarray}
Substituting these values into expressions (\ref{w5}) and (\ref{w7}), we can 
calculate the curvatures for parallel and anti-parallel spins as  
  \begin{eqnarray}
             && 
R_{Free/Bound-Free/Bound}^{anti-parallel}=l_{1}^{1-1}+l_{1}^{1-2}+l_{1}^{1-3}
+ 
             (l'_{1})^{1-1}+(l'_{1})^{1-2}+(l'_{1})^{1-3}
=0\nonumber\\
&& R_{Free/Bound-Free/Bound}^{
parallel}=l_{2}^{1-1}+l_{2}^{1-2}+l_{2}^{1-3}- 
    (l'_{2})^{1-1}-(l'_{2})^{1-2}-(l'_{2})^{1-3}= 0,\label{w19}
  \end{eqnarray}
and then inserting into (\ref{w11}) for the current density we obtain 
\begin{eqnarray}
    && J \approx  0.
    \label{w20}
  \end{eqnarray}
Hence, we conclude that for the  standard  graphene without defects, the current density 
is 
zero and thus the electrons do not collectively move in any specific direction. 
Consequently, electron moves randomly and superconductivity disappears. 


\subsection{Graphene with heptagonal defects}

\begin{figure}[ht]
\centering
\includegraphics[width=0.35\linewidth]{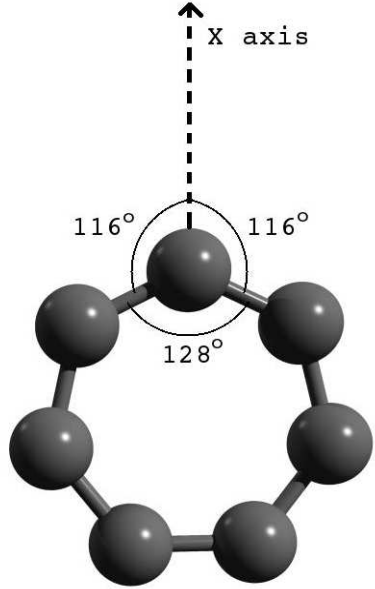}
\caption{{\it{The angle between electron pairs in the graphene with a heptagonal 
defect.}}}
\label{heptagonal}
\end{figure}

Let us now investigate the case of graphene with heptagonal defects. Similar to hexagonal 
case, we assume one axis along $X$ (see Fig. \ref{heptagonal}), and we consider that the 
couplings for free and bound electrons are the same. We calculate the couplings of 
parallel and anti-parallel spins in terms of their angles with respect to the $X$-axis as:
\begin{eqnarray}
  &&\!\!\!\!\!\!\!\!\!\!\!\!\!\!\!\!\!\!\!\!\!\!\!
   l_{1}^{1-1}\approx \cos(0)=1,
   \quad l_{1}^{1-2}\approx \cos(116)=-0.434, 
\quad l_{1}^{
1-3}\approx \cos(244)=-0.434,
\nonumber\\
  &&\!\!\!\!\!\!\!\!\!\!\!\!\!\!\!\!\!\!\!\!\!\!\!
(l'_{1})^{1-1}\approx \sin(0)=0,
\quad 
(l'_{1})^{1-2}\approx \sin(116)=0.901,
\quad (l'_{1})^{1-3}\approx 
\sin(244)=-0.901,
 \end{eqnarray}
          and
 \begin{eqnarray}
  &&\!\!\!\!\!\!\!\!\!\!\!\!\!\!\!\!\!\!\!\!\!\!\!
   l_{2}
^{1-1}\approx \cos(60)=\frac{1}{2},
\quad l_{2}^{1-2}\approx \cos(176)=-0.997,
\quad 
l_{2}^{1-3}\approx \cos(304)=0.563,
\nonumber\\
  &&\!\!\!\!\!\!\!\!\!\!\!\!\!\!\!\!\!\!\!\!\!\!\!
  (l'_{2})^{1-1}\approx \sin(60)=0.866, 
\quad (l'_{2})^{1-2}\approx \sin(116)=0.075, \quad (l'_{2})^{1-3}\approx 
\sin(304)=-0.826.
\label{w21}
  \end{eqnarray}
Substituting these values into  (\ref{w5}) and (\ref{w7}), we can 
calculate the curvatures for parallel and anti-parallel spins as  
   \begin{eqnarray}
   && \!\!\!\!\!\!\!\!\!\!\!\!\!\!\!\!\!\!
R_{Free/Bound-Free/Bound}^{anti-parallel}=l_{1}^{1-1}+l_{1}^{1-2}+l_{1}^{1-3}
+                  
(l'_{1})^{1-1}+(l'_{1})^{1-2}+(l'_{1})^{1-3}= 0.132
 \nonumber
 \\
 && \!\!\!\!\!\!\!\!\!\!\!\!\!\!\!\!\!\!
R_{Free/Bound-Free/Bound}^{parallel}=l_{2}^{1-1}+l_{2}^{1-2}+l_
{2}^{1-3}-               
(l'_{2})^{1-1}-(l'_{2})^{1-2}-(l'_{2})^{1-3}= -0.048.
\label{w22}
                      \end{eqnarray}
Note that the positive sign for anti-parallel spins implies that the electrons 
are attracted by electrons of neighbor molecules and that the coupling of anti-parallel 
spins is along the $X$-axis, while the negative sign for parallel spins means that 
electrons are repelled by parallel spins in neighbor molecules and that the coupling of 
parallel spins is along the negative $X$-axis. Substituting the values (\ref{w22}) into  
the 
current density (\ref{w11}),  we obtain            
\begin{equation}
J \approx  0.458.
\label{w23}
\end{equation}
Since the current density is positive,  we deduce that electrons are repelled by neighbor 
molecules and move along the $X$-axis (the curvature produced by parallel spins is larger 
than the curvature produced by anti-parallel spins and therefore a negative force is 
applied to electrons and they move in opposite directions with respect to the molecule). 
This result is also in agreement with previous predictions that the curvature of 
heptagonal defect is negative \cite{B1, B14a,Gonzalez:2009je,B14b}.
                        
\subsection{Graphene with pentagonal defects}

\begin{figure}[ht]
\centering
\includegraphics[width=0.35\linewidth]{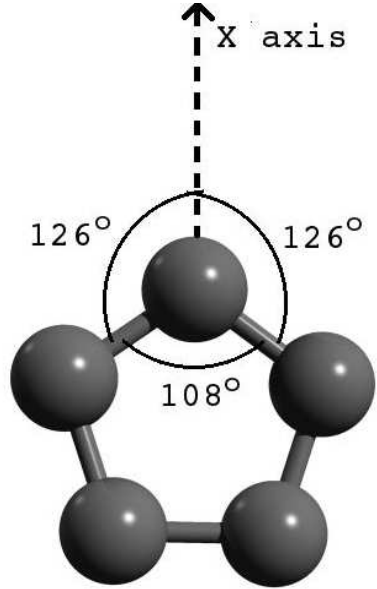}
\caption{{\it{The angle between electron pairs in the graphene with a pentagonal 
defect.}}}
\label{pentagonal}
\end{figure}

\begin{figure}[ht]
\centering
\includegraphics[width=0.7\linewidth]{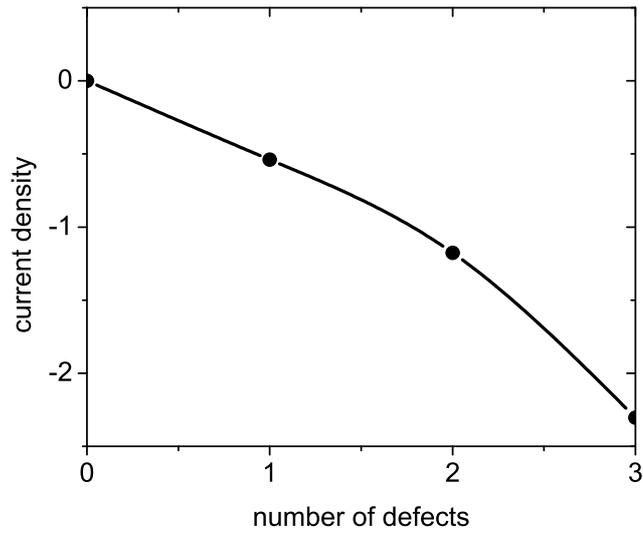}
\caption{{\it{The dependence of the current density on the number of 
pentagonal defects.}}}
\label{FigP}
\end{figure}

Finally, let us investigate the case of pentagonal defects. Similarly to the previous 
cases, we consider one axis along $X$ (see Fig. \ref{pentagonal}), and we assume that the
couplings for free and bound electrons are the same. We calculate the couplings of 
parallel and anti-parallel spins in terms of their angles with respect to 
the $X$-axis as:
\begin{eqnarray}
 && \!\!\!\!\!\!\!\!\!\!\!\!\!\!\!\!\!\!
 l_{1}^{1-1}\approx \cos(0)=1, \quad 
l_{1}^{1-2}\approx \cos(126)=-0.587, 
\quad l_{1}^{1-3}\approx \cos(234)=-0.587,
\nonumber\\
 && \!\!\!\!\!\!\!\!\!\!\!\!\!\!\!\!\!\!
(l'_{1})^{1-1}\approx \sin(0)=0, 
\quad 
(l'_{1})^{1-2}\approx \sin(126)=0.809,
\quad (l'_{1})^{1-3}
\approx 
\sin(234)=-0.809,
 \end{eqnarray}
\begin{eqnarray}
 && \!\!\!\!\!\!\!\!\!\!\!\!\!\!\!\!\!\!\!\!
 l_{2}
^{1-1}\approx \cos(60)=\frac{1}{2},  \quad l_{2}^{1-2}\approx \cos(186)=-0.994, \quad 
l_{2}^{1-3}\approx \cos(294)=0.406,
\nonumber\\
 && \!\!\!\!\!\!\!\!\!\!\!\!\!\!\!\!\!\!\!\!
 (l'_{2})^{1-1}\approx \sin(60)=0.866,  
\  (l'_{2})^{1-2}\approx \sin(186)=-0.104,
\  (l'_{2})^{1-3}\approx 
\sin(294)=-0.913.
\label{w24}
\end{eqnarray}
Substituting these values into expressions (\ref{w5}) and (\ref{w7}), we can 
calculate the curvatures for parallel and anti-parallel spins as  
\begin{eqnarray}
 && \!\!\!\!\!\!\!\!\!\!\!\!\!\!\!\!\!\!
R_{Free/Bound-Free/Bound}^{anti-parallel}=l_{1}^{1-1}+l_{1}^{1-2}+l_{1}^{
1-3}+   (l'_{1})^{1-1}+(l'_{1})^{1-2}+(l'_{1})^{1-3}=  -0.174 
\nonumber\\
 && \!\!\!\!\!\!\!\!\!\!\!\!\!\!\!\!\!\!
R_{Free/Bound-Free/Bound}^{parallel}=l_{2}^{1-1}+l_{2}
^{1-2}+l_{2}^{1-3}-                                    
(l'_{2})^{1-1}-(l'_{2})^{1-2}-(l'_{2})^{1-3}=0.063.
\label{w25}
   \end{eqnarray}
Substituting these values into the current density (\ref{w11}) we obtain          
\begin{equation}
J \approx  -0.539.\label{w26}
\end{equation}
The negative value of the current density implies that the electrons are absorbed by 
pentagonal defects and move along the negative $X$-axis, i.e  this type of defects 
induces 
a force to the free electrons and leads them to move towards the molecule. Hence, 
increasing the number of defects, the current density increases and the 
graphene tends to be a superconductor. In Fig. \ref{FigP}, we depict the dependence of 
the current density on different number of pentagonal defects.
            
In summary, in this section we found that heptagonal defects repel electrons while 
pentagonal defects absorb electrons. This is a consequence of the fact that  in 
heptagonal 
defects the curvature of parallel spins is larger than the curvature produced by 
anti-parallel spins, while, in pentagonal defects,  the curvature produced by parallel 
spins 
is smaller than the curvature produced by anti-parallel spins. These results are in 
agreement with previous analyses for defects in the graphene  \cite{B1, 
B14a,Gonzalez:2009je,B14b}.
        

\section{The current density in a graphene wormhole}\label{o3}

\begin{figure}[ht]
\centering
\includegraphics[width=0.7\linewidth]{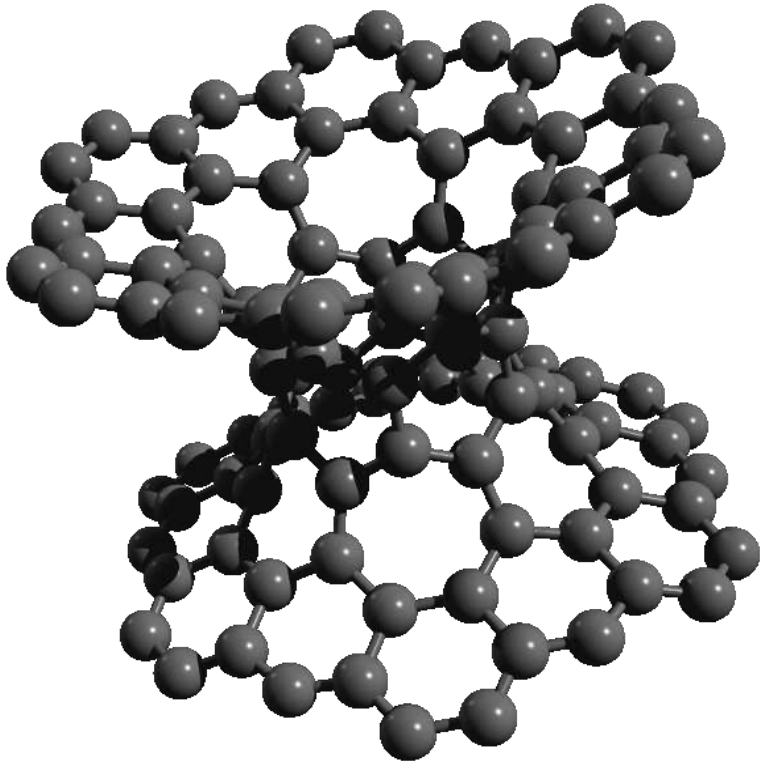}
\caption{{\it{Graphene wormhole with 12 heptagonal defects.}}}
\label{wormhole2}
\end{figure}

\begin{figure}[ht]
\centering
\includegraphics[width=0.7\linewidth]{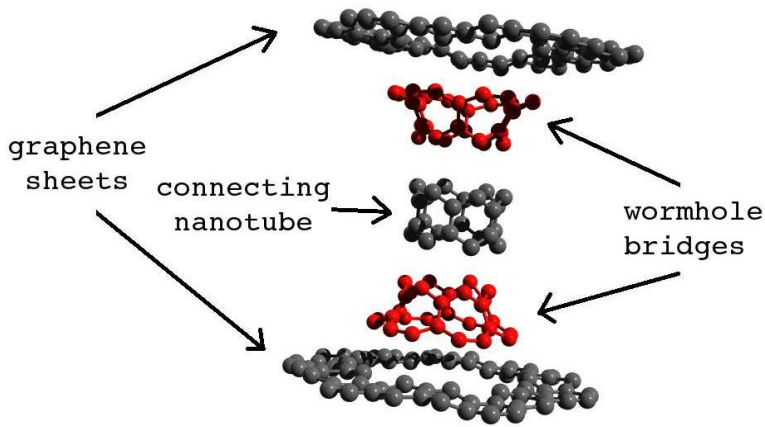}
\caption{{\it{A schematic split of a graphene wormhole with 12 heptagonal defects.}}}
\label{wormhole2a}
\end{figure}

Let us consider the formalism and methods of the 
previous section in the case of a graphene wormhole. Such an object is created when the 
structure of the plain graphene is disrupted by the presence of heptagonal 
defects \cite{ B14a,Gonzalez:2009je,B14b}.
In the following, without loss of generality, we take into account the case of 12 defects, 
and in 
Fig. \ref{wormhole2},  we give a picture of such a graphene wormhole.
The graphene wormhole consists of the upper and lower graphene sheets, each one connected 
to a wormhole bridge which is connected to the common connecting nanotube, as can be seen 
in Fig. \ref{wormhole2a}. 

Let us now calculate the current density for this case. According to the results of the 
previous section, the current density should be zero in both graphene sheets and positive 
($J \approx  0.458$) in the case of a heptagonal defect. Additionally, we expect that,  
in 
the place of the wormhole bridges  where 6 defects appear (and hence 12 in 
total,  if we consider both sides of the connecting nanotube), the overall current 
density 
should be proportional to the current density for 1 defect, namely  
\begin{equation}
  J \approx  6\cdot 0.458=2.748.
\end{equation}
The crucial question from the experimental point of view is what value of the current 
density would be measured in the region around the middle of the connecting nanotube. One 
would expect that exactly at the middle, the current densities coming from opposite 
directions eliminate each other, and thus the overall current density is zero. 
On the other hand, at small distances away from the middle, the geometry of the 
structure differs significantly from the geometry in the wormhole bridge, and, therefore, 
one should expect a change of the current density. Hence, we can apply the procedure of 
the previous section in order to explicitly calculate the exact value.
 
In order to perform the calculation, we need the appropriate value of the angle 
$\theta_1$, substitute  it into the expressions (\ref{w18a}),(\ref{w18}) for the 
couplings 
$l_1$,$l_2$,  and calculate the curvatures of the parallel and anti-parallel spins using 
(\ref{w5}) and (\ref{w7}). Finally, the current density is calculated using (\ref{w11}). 

\begin{figure}[ht]
\centering
\includegraphics[width=0.9\linewidth]{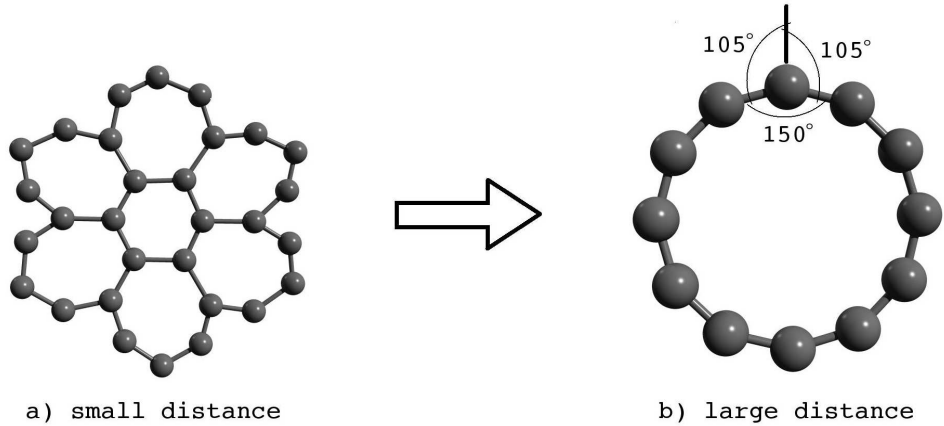}
\caption{{\it{The geometric structure at small and large distance  from the defects.}}}
\label{GW2}
\end{figure}

The value of $\theta_1$ depends on the distance from the defects: at shorter distances 
the character of the physical quantities is either the same as in the case of a simple 
defect, or it is the superposition of $n$ defects, respectively. At larger distances from 
the place of $n$ heptagonal defects, the geometry is approaching the case of the 
structure with 1 defect consisting of $6+n$ vortices, as can be seen from Fig. \ref{GW2}.
Hence, we deduce that at large distances from the wormhole bridges, the value of the 
current density corresponds to the case of a dodecagon ($12=6+n$, where $n=6$ for each of 
the two parts of the wormhole structure). In this case, the value of $\theta_1$ 
corresponds to Fig. \ref{GW2}b, namely $\theta_1=105$. Thus, (\ref{w18a}),(\ref{w18}) 
lead to
\begin{eqnarray}
 && \!\!\!\!\!\!\!\!\!\!\!\!\!\!\!\!\!\!
 l_{1}^{1-1}\approx \cos(0)=1,
\quad l_{1}^{1-2}\approx \cos(105)=-0.259, \quad 
l_{1}^{1-3}\approx \cos(255)=-0.259,
\nonumber\\
 && \!\!\!\!\!\!\!\!\!\!\!\!\!\!\!\!\!\!
 (l'_{1})^{1-1}\approx \sin(0)=0,
\quad (l'_{1})^{1-2}\approx \sin(105)=0.966,
\quad 
(l'_{1})^{1-3}\approx \sin(255)=-0.966,
\end{eqnarray}
and
\begin{eqnarray}
 && \!\!\!\!\!\!\!\!\!\!\!\!\!\!\!\!\!
 l_{2}^{1-1}\approx 
\cos(60)=\frac{1}{2},  \quad l_
{2}^{1-2}\approx \cos(165)=-0.966, \quad l_{2}^{1-3}\approx \cos(315)=0.707,
\nonumber\\
 && \!\!\!\!\!\!\!\!\!\!\!\!\!\!\!\!\!
 (l'_{2})^{1-1}\approx \sin(60)=\frac{\sqrt{3}}{2},
\quad (l'_{2})^{1-2}\approx 
\sin(165)=0.259, \quad (l'_{2})^{1-3}\approx \sin(315)=-0.707.
\label{GW3}
\end{eqnarray}
Substituting these values into expressions (\ref{w5}) and (\ref{w7}), we can 
calculate the curvatures for parallel and anti-parallel spins as  
\begin{eqnarray}
&& 
R_{Free/Bound-Free/Bound}^{anti-parallel}=0.482,\nonumber\\
&&R_{Free/Bound-Free/Bound}^{
parallel}
=-0.177,\label{GW4}
\end{eqnarray}
and thus for the current density (\ref{w11}) we obtain  
\begin{eqnarray}
&& J \approx  1.960.
\label{GW5}
\end{eqnarray}

Hence, we deduce that in the graphene wormhole, the current density is zero in the upper 
and lower graphene sheet, then it rises up to value $2.748$ in the wormhole bridges, 
and in the connecting nanotube it decreases to the value $1.960$. In the centre of the 
connecting nanotube, the current density is exactly zero due to the mutual 
elimination.

\begin{figure}[ht]
\centering
\includegraphics[width=0.7\linewidth]{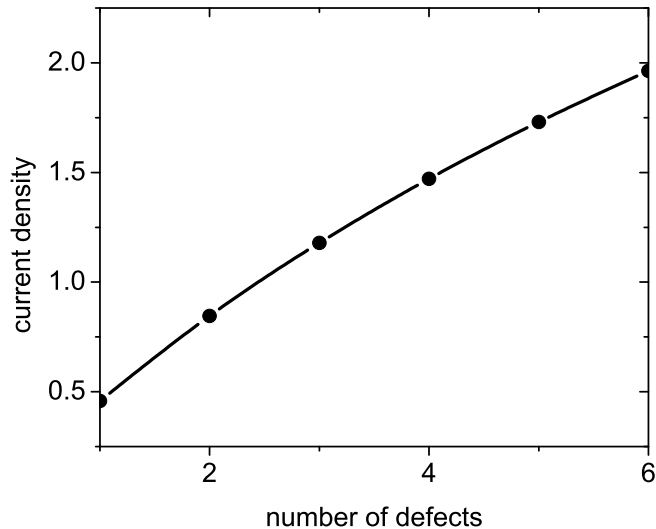}
\caption{{\it{The dependence of the current density, close to the middle of the 
connecting 
nanotube, on the number of defects. }}}
\label{GW6}
\end{figure}

Let us  close this section by mentioning that the number of defects in the graphene 
wormhole can differ from 2 to 12 defects, i.e. from 1 to 6 defects at each side. 
Therefore, the upper and lower sheets will not have the geometry of the plain graphene, 
and the corresponding current density will change, as well as the current density close 
to 
the middle of the connecting nanotube. In Fig. \ref{GW6} we depict the dependence 
of the current density close to the middle of the connecting 
nanotube, on the number of defects. 


\section{Summary and Discussion}
\label{sum}

In this work we have studied, in the context of graphene structure, the exchange of gauge 
fields 
between electrons, which can 
be considered as a sort of effective gravitons. Such exchange  leads  to the emergence of conductivity. In 
this way, three types of curvature are produced, one between free electrons, one between 
bound electrons, and one between free and bound electrons. These fields create a sort of 
an 
effective gravity 
with positive curvature between parallel spins, and anti-gravity with negative curvature 
between parallel spins. The current density of free electrons has been obtained in terms 
of inequality between curvatures of parallel spins and those between anti-parallel spin

In particular, in Sec. 2, we  obtained the current density  in terms of curvature
 of parallel spins and anti-parallel spins and in Sec. 3 and 4, by substituting the
relations between 
curvatures and parameters of graphene molecule, we have calculated the 
current 
density in terms of the angle between atoms with respect to the center of graphene molecule. In 
fact, in 
Sec. 2, using the concepts in M-theory, we introduced an action for conductivity 
in a 
graphene system in terms of gauge fields and fermions. Then, we  obtained the relation 
between 
gauge fields and curvature of parallel spins and anti-parallel spins.  Also, it has been 
shown that 
the curvature which is produced between parallel spins has an opposite sign with respect to the 
curvature 
which is generated  between anti-parallel spins. These curvatures have a direct relation with the 
effective energy-momentum tensors. Also, changes in momentums  have a direct relation with the applied 
force between 
spinors.  Consequently, the force between parallel spin has an opposite sign with respect to 
the force 
between anti-parallel spins as it is possible to  observe in laboratory experiments. Using the relations between gauge 
fields, 
spinors and curvature, it is possible to  obtain the energy of system in terms of difference 
between 
curvatures of parallel spins and anti-parallel spins. Besides, using this 
energy, the current density in terms of curvatures has been derived.

These results can be applied to realistic  graphene structures and a relation 
between 
curvatures and graphene parameters,  like the angles between atoms with respect to the molecular center, 
can be obtained.  Using these relations, the current density, in terms of 
such angles, can be calculated. We have shown that, for the standard 
graphene 
with hexagonal molecules, the current density is zero and thus the electrons do not 
collectively 
move in any given direction. Consequently, electrons move randomly and 
superconductivity 
disappears. For graphene with heptagonal defects, since the current density is positive, 
it is possible to  deduce 
that electrons are repelled by neighbor
 molecules and move outward from molecules (the curvature produced by parallel spins is 
larger
 than the curvature produced by anti-parallel spins and therefore a negative force is 
applied
 to electrons. As consequence,   they move in opposite directions with respect to the molecule). This 
result is 
also in agreement with previous predictions that the curvature of 
 heptagonal defect is negative \cite{B1, B14a,Gonzalez:2009je,B14b}. For graphene with 
pentagonal 
defects, the negative value of the current density implies that the electrons are 
absorbed by 
 pentagonal defects and move towards molecules, i.e  this type of defects induces 
 a force to the free electrons and leads them to move towards the molecules. Hence, 
 increasing the number of defects, the current density increases and the 
 graphene tends to be a superconductor. 
 
 In Sec. 4,  we  deduced that, in the 
graphene 
wormhole, the current density is zero in the upper 
 and lower graphene sheet, then it rises up to positive value in the wormhole bridges. On the other hand,  
  in the connecting nanotube, it decreases to the lower values. In the centre of the 
 connecting nanotube, the current density is exactly zero due to the mutual 
 elimination.

In standard graphene, due to its symmetry, the curvature of anti-parallel spins is 
canceled 
by the curvature of parallel spins,  and hence the total current density of free 
electrons becomes zero. Thus, the free electrons do not move in any special direction and 
therefore conductivity disappears. On the other hand, for some particular types of 
defects in the graphene, anti-parallel spins come closer mutually, their curvature 
increases, and a sort of modified gravity emerges. Consequently, the current density grows 
and 
conductivity increases. Similarly, for some other types of defects, parallel 
spins approach each other, their negative curvature increases and (modified) anti-gravity 
appears. In this case,  the sign of current density reverses, the electrons move in 
opposite direction, and a new conductivity appears along this new direction. 

In the case of more complicated structures, such as graphene wormholes, the current  
density in the different regions of the molecular surface has different values. Moreover, 
the curvature induced by the defects cannot be clearly determined from the type of the 
defects and it is given by the chemical structure of the whole molecule. As a result, 
the current density arising from these defects depends on the distance from them. We 
derived the current density for the graphene wormhole which includes 12 heptagonal 
defects, and we have found an approximation for the cases where the number of defects 
changes. The current density on the curved wormhole sheets in such modified structures 
remains an open question.

Furthermore, it is possible to  show that each defect produces a given form of extended  gravity. 
In particular, as reported in Appendix C, no defect, heptagonal defects and pentagonal defects give rise to 
different forms of $f(R)$ gravity. 

In general,  the type of defects can be determined by the chemical 
structure 
of the whole molecule. This means that  each  defect produces a different type of gravity and 
current  density. For example, pentagonal molecules absorb electrons, while heptagonal molecules 
repel them.  By inserting defects in graphene suitable places, electrons are repelled by some molecules 
and 
absorbed by other ones and move in an given direction. This helps us to design a 
graphene which  conducts electrons in an given  direction. In fact, for producing a good superconductor, 
it is  needed to insert  heptagonal and pentagonal molecules in suitable places between hexagonal 
graphene molecules.  On the other hand, this means that the gravitational anologue picture  can discriminate among the various graphene structures.

As final remark, it is worth noticing that, from an experimental point of view, these 
graphene defects structures  could be probed using currents facilities and experimental 
apparatuses working in several laboratories \cite{exp1,exp2} considering that the presence 
of the absence of precise current densities are the final test beds for the models. On the 
other hand, analogue models of gravity \cite{analogue}  could be realized by graphene 
considering the 2-form tensor fields discussed above. In a forthcoming paper, we will 
discuss in details possible experimental realizations  of these structures.

\begin{acknowledgments}
The work by A. Sepehri has been financially supported by the Research Institute 
for Astronomy and Astrophysics of Maragha (RIAAM), Iran, under research project 
No.1/4165-14. The work was partly supported by VEGA Grant No. 2/0009/16. R. Pincak would 
like to thank the TH division in CERN for hospitality. This work was partially  supported 
by the JSPS Grant-in-Aid for Young Scientists (B) \# 25800136 and the research-funds 
given by  Fukushima University (K.B.). S. Capozziello acknowledges financial support of 
INFN (iniziative specifiche TEONGRAV and QGSKY). This article is also based upon work from 
COST action CA15117 (CANTATA), supported by COST (European Cooperation in Science and 
Technology). 
\end{acknowledgments}

\begin{appendix}

\section{The   action components  (\ref{w1}) in 
terms of couplings of parallel and anti-parallel spins} 
\label{Appterms}

In this Appendix we use the  definitions (\ref{w4}) in order to calculate the 
different terms of (\ref{w1}) in terms of couplings of parallel and anti-parallel spins 
\cite{Sepehri:2016sjq}. In particular, we obtain: 
\begin{eqnarray}
 &&\!\!\!\!\!\!\!\!\!\!\!\!\!\!\!\!\!\!\!\!\!\!\!\!\!\!\!\!\!\!\!
 \langle
 F^{abc},F_{abc}\rangle_{Free-Free} = 
A^{ab}i\sigma_{ij}^{2}\partial_{a}^{i}\psi^{j}_{b}+\sigma^{0}_{ij}\psi^{
\dag a,i}\psi^{j}_{a}-\sigma^{1}_{ij}\psi^{\dag a, i}\psi^{j}_{a}
\nonumber \\
 &&\ \ \ \ \ \ \ \ \ \ \ \ \ \ \ 
 +
 \sigma^{0}_{i'i}(\psi^{\dag a, 
i'}i\sigma^{0}_{i'j}\sigma^{1}_{jk}\partial^{a,j}\psi^{k}_{a})(\psi^{\dag  
i}_{a}i\sigma^{0}_{ij}\sigma^{1}_{jk}\partial^{a,j}\psi^{k}_{a})
\nonumber \\
 &&\ \ \ \ \ \ \ \ \ \ \ \ \ \ \ 
 +
 \sigma^{1}_{i'i}(\psi^{\dag a, 
i'}i\sigma^{0}_{i'j}\sigma^{1}_{jk}\partial^{a,j}\psi^{k}_{a})(\psi^{\dag  
i}_{a}i\sigma^{0}_{ij}\sigma^{1}_{jk}\partial^{a,j}\psi^{k}_{a})
\nonumber \\
 &&\ \ \ \ \ \ \ \ \ \ \ \ \ \ \ 
 -
 \sigma^{0}_{i'i}(\psi^{\dag a, 
i'}i\sigma^{1}_{i'j}\sigma^{1}_{jk}\partial^{a,j}\psi^{k}_{a})(\psi^{\dag  
i}_{a}i\sigma^{1}_{ij}\sigma^{1}_{jk}\partial^{a,j}\psi^{k}_{a}),
 \end{eqnarray}
  and
  \begin{eqnarray}
 &&\!\!\!\!\!\!\!\!\!\!\!\!\!
 \langle 
\partial^{b}\partial^{a}X^{i},\partial_{b}\partial_{a}X^{i}\rangle=\varepsilon^{abc}
\varepsilon^{ade}(\partial_{b}\partial_{c}X^{i}_{\alpha})(\partial_{e}\partial_{d}X^{i}_{ 
\beta})=\nonumber \\
 &&\ \ \ \ \ \ \ \ \ \ \ \ \ \ \ \ \ \ \ \     
 \Psi^{\dag a,U}\langle
     F_{abc},F^{a'bc}\rangle\Psi_{a'}^{L}+\Psi^{\dag a,L}\langle
     F_{abc},F^{a'bc}\rangle\Psi_{a'}^{U}- \Psi^{\dag a,U}\langle
     F_{abc},F^{a'bc}\rangle\Psi_{a'}^{U}
     \nonumber \\
 &&\ \ \ \ \ \ \ \ \ \ \ \ \ \ \ \ \ \ \ \ 
 -\Psi^{\dag a,L}\langle
     F_{abc},F^{a'bc}\rangle\Psi_{a'}^{L}
     +
     \Psi^{\dag a,L}\Psi^{\dag d,U}\partial_{d}\partial^{d'}\langle
     F_{abc},F^{a'bc}\rangle\Psi_{a'}^{L}\Psi^{U}_{d'}\nonumber \\
 &&\ \ \ \ \ \ \ \ \ \ \ \ \ \ \ \ \ \ \ \ 
     -
     \Psi^{\dag a,L}\Psi^{\dag d,U}\partial_{d}\langle
     F_{abc},F^{a'bc}\rangle\Psi_{a'}^{L}
     -
     \Psi^{\dag a,U}\Psi^{\dag d,L}\partial_{d}\langle
     F_{abc},F^{a'bc}\rangle\Psi_{a'}^{U} 
 \nonumber \\
            &&\ \ \ \ \ \ \ \ \ \ \ \ \ \ \ \ \ \ \ \ 
           + \psi^{\dag i,U}\langle
               F_{ijk},F^{i'jk}\rangle\psi_{i'}^{L}+\psi^{\dag i,L}\langle
               F_{ijk},F^{i'jk}\rangle\psi_{i'}^{U}-
               \psi^{\dag i,U}\langle
               F_{ijk},F^{i'jk}\rangle\psi_{i'}^{U}
               \nonumber \\
 &&\ \ \ \ \ \ \ \ \ \ \ \ \ \ \ \ \ \ \ \ 
 -\psi^{\dag i,L}\langle
               F_{ijk},F^{i'jk}\rangle\psi_{i'}^{L}+
               \psi^{\dag i,L}\psi^{\dag m,U}\partial_{m}\partial^{m'}\langle
               F_{ijk},F^{i'jk}\rangle\psi_{i'}^{L}\psi^{U}_{m'}
               \nonumber \\
 &&\ \ \ \ \ \ \ \ \ \ \ \ \ \ \ \ \ \ \ \ 
             -  \psi^{\dag i,L}\psi^{\dag m,U}\partial_{m}\langle
               F_{ijk},F^{i'jk}\rangle\psi_{i'}^{L}- \psi^{\dag i,U}\psi^{\dag 
m,L}\partial_{m}\langle
               F_{ijk},F^{i'jk}\rangle\psi_{i'}^{U} 
               \nonumber \\
 &&\ \ \ \ \ \ \ \ \ \ \ \ \ \ \ \ \ \ \ \  +    \Psi^{\dag a,U}\langle
                         F_{abc},F^{i'bc}\rangle\psi_{i'}^{L}+\Psi^{\dag a,L}\langle
                         F_{abc},F^{i'bc}\rangle\psi_{i'}^{U}-
                         \Psi^{\dag a,U}\langle
                         F_{abc},F^{i'bc}\rangle\psi_{i'}^{U}
                         \nonumber \\
 &&\ \ \ \ \ \ \ \ \ \ \ \ \ \ \ \ \ \ \ \  -\Psi^{\dag a,L}\langle
                         F_{abc},F^{i'bc}\rangle\psi_{i'}^{L}+
                         \Psi^{\dag a,L}\Psi^{\dag d,U}\partial_{d}\partial^{i'}\langle
                         F_{abc},F^{j'bc}\rangle\psi_{j'}^{L}\psi^{U}_{i'} 
                         \nonumber \\
 &&\ \ \ \ \ \ \ \ \ \ \ \ \ \ \ \ \ \ \ \ 
                         -\Psi^{\dag a,L}\Psi^{\dag d,U}\partial_{d}\langle
                         F_{abc},F^{i'bc}\rangle\psi_{i'}^{L}- 
                         \Psi^{\dag a,U}\Psi^{\dag 
d,L}\partial_{d}\langle
                         F_{abc},F^{i'bc}\rangle\psi_{i'}^{U}.
 \end{eqnarray}
 Additionally, since $(\psi^{U}_{a})^{2}=0$ and $(\psi^{L}_{a})^{2}=0$, we get
  \begin{eqnarray}
&&
F(X)=\Sigma_{j}X_{j}^{2}=\Sigma_{j}[\psi^{
U}_{a}A^{ab,j}\psi^{L}_{b}-\psi^{L}_{a}A_{ab,j}\psi^{U}_{b}]^{2}
\nonumber \\ 
 &&\ \ \ \ \ \ \ \ \, 
=\Sigma_{j}(\psi^{U}_{a})^{2}(A^{ab,j})^{2}(\psi^{L}_{b})^{2}+(\psi^{L}_{a})^{2}(A_{ab,j}
)^{2}(\psi^{U}_{b})^{2}
\nonumber \\
 &&\ \ \ \ \ \ \ \ \ \ \  \
 -(\psi^{U}_{a}A^{ab,j}\psi^{L}_{b})(\psi^{L}_{a}A_{ab,j}\psi^{U}_{b
} )
 -(\psi^{L}_{a}A_{ab,j}\psi^{U}_{b})(\psi^{U}_{a}A^{ab,j}\psi^{L}_{b})=0
  \end{eqnarray}
  and
    \begin{equation}
\langle[X^{k},X^{i},X^{j}],[X_{k},X_{i},X_
{j}]\rangle=\Sigma_{n}\Sigma_{m}\alpha_{n+m}(\psi^{L})^{2n}(\psi^{U})^{2m}=0.\label{w6}
 \end{equation}
 In the above expressions $a,b,c$ are indices of bound electrons, while $i,j,k$ are 
indices of free electrons, and $U,L$ refers to upper and lower spins. Additionally,    
 we have used the Pauli matrices definition as: $\sigma_{ij}^{1}=\Big{(}\begin{array}{cc}
0 & 1 \\
1 & 0
\end{array}\Big{)} $, $\sigma_{ij}^{0}=\Big{(}\begin{array}{cc}
1 & 0 \\
0 & 1
\end{array}\Big{)} $, $\sigma_{ij}^{2}=\Big{(}\begin{array}{cc}
 & -i \\
i & 0
\end{array}\Big{)} $.

 \section{ Expression of $\langle 
\partial^{b}\partial^{a}X^{i},\partial_{b}\partial_{a}X^{i}\rangle$ in terms of 
curvatures }
 \label{Bigfermcurvterm}

The term $\langle 
\partial^{b}\partial^{a}X^{i},\partial_{b}\partial_{a}X^{i}\rangle$ is expressed in terms 
of curvatures as:
 \begin{eqnarray}
 &&\!\!\!\!\!\!\!\!\!\!\!\!\!\!\!\!\!\!\!\!\!\!\!\! 
 \langle \partial^{b}\partial^{a}X^{i},\partial_{b}\partial_{a}X^{i}\rangle = 
 \Psi^{\dag a,U}
R_{aa'}^{anti-parallel} \Psi^{a',L}+\Psi^{\dag 
a,L}R_{aa'}^{anti-parallel}\Psi^{a',U}
\nonumber \\
&& \ \ \ \ \ \ \ \ \ \ \ \ \ \ \,
     - \Psi^{\dag a,U}R_{aa'}^{parallel}\Psi^{a',U}-\Psi^{\dag a,L}R_{aa'}^{parallel} 
\Psi^{a',L}
\nonumber \\
&& \ \ \ \ \ \ \ \ \ \ \ \ \ \ \, +\Psi^{\dag a,L}\Psi^{\dag 
d,U}\partial_{d}\partial^{d'}(R_{aa'}^{parallel}+R_{aa'}^{anti-
parallel})\Psi^{a',L}\Psi^{U}_{d'}
\nonumber \\
&& \ \ \ \ \ \ \ \ \ \ \ \ \ \ \, 
     - \Psi^{\dag a,L}\Psi^{\dag 
d,U}\partial_{d}(R_{aa'}^{parallel}+R_{aa'}^{anti-parallel}) \Psi^{a',L}
\nonumber \\
&& \ \ \ \ \ \ \ \ \ \ \ \ \ \ \, -\Psi^{\dag a,U}\Psi^{\dag 
d,L}\partial_{d}(R_{aa'}^{parallel}+R_{aa'}^{anti-parallel})\Psi^{
a',U} 
\nonumber \\
&& \ \ \ \ \ \ \ \ \ \ \ \ \ \ \,
+  \psi^{\dag i,U}R_{ii'}^{anti-parallel} 
\psi^{i',L}+\psi^{\dag 
i,L}R_{ii'}^{
anti-parallel}\psi^{i',U}
\nonumber \\
&& \ \ \ \ \ \ \ \ \ \ \ \ \ \ \, -\psi^{\dag 
i,U}R_{ii'}^{parallel}\psi^{i',U}-\psi^{\dag 
i,L}R_{ii'}^{
parallel}\psi^{i',L}
\nonumber \\
&& \ \ \ \ \ \ \ \ \ \ \ \ \ \ \,
                +
                \psi^{\dag i,L}\psi^{\dag 
m,U}\partial_{m}\partial^{m'}(R_{ij'}^{parallel}+R_{ij'}
^{anti-parallel})\psi^{i',L}\psi^{U}_{m'}
\nonumber \\
&& \ \ \ \ \ \ \ \ \ \ \ \ \ \ \, -
                \psi^{\dag i,L}\psi^{\dag 
m,U}\partial_{m}(R_{ij'}^{parallel}+R_{ij'}^{anti-
parallel})\psi^{i',L}
\nonumber \\
&& \ \ \ \ \ \ \ \ \ \ \ \ \ \ \,
                -
                \psi^{\dag i,U}\psi^{\dag 
m,L}\partial_{m}(R_{ij'}^{parallel}+R_{ij'}^{anti-
parallel})\psi^{i',U} 
\nonumber \\
&& \ \ \ \ \ \ \ \ \ \ \ \ \ \ \,
+
                      \Psi^{\dag 
a,U}R_{ai'}^{anti-parallel}\psi^{i',L}+\Psi^{\dag 
a,L}R_{
ai'}^{anti-parallel} \psi^{i',U}
\nonumber \\
&& \ \ \ \ \ \ \ \ \ \ \ \ \ \ \, -
                          \Psi^{\dag a,U}R_{ai'}^{parallel}\psi^{i',U}-\Psi^{\dag 
a,L}R_{ai'}^{
parallel}\psi^{i',L}
\nonumber \\
&& \ \ \ \ \ \ \ \ \ \ \ \ \ \ \, +
                          \Psi^{\dag a,L}\Psi^{\dag 
d,U}\partial_{d}\partial^{i'}(R_{ai'}^{
parallel}+R_{ai'}^{anti-parallel})\psi_{j'}^{L}\psi^{i',U}
\nonumber \\
  && \ \ \ \ \ \ \ \ \ \ \ \ \ \ \, -
         \Psi^{\dag a,L}\Psi^{\dag 
d,U}\partial_{d}(R_{ai'}^{parallel}+R_{ai'}^{
anti-parallel})\psi^{i',L}
\nonumber \\
&& \ \ \ \ \ \ \ \ \ \ \ \ \ \ \, -
     \Psi^{\dag a,U}\Psi^{\dag 
d,L}\partial_{d}(R_{ai'}^{parallel}+R_{ai'}^{
anti-parallel})\psi^{i',U}
\nonumber \\
&& \ \ \ \ \ \ \ \ \ \ \ 
  \approx
  ( R_{Free-Free}^{parallel})^{2}+( 
R_{Free-
Free}^{anti-parallel})^{2}+( R_{Free-Bound}^{parallel})^{2}
\nonumber \\
&& \ \ \ \ \ \ \ \ \ \ \ \ \ \ \,  +
(R_{Free-Bound}^{anti-
parallel})^{2}+( R_{Bound-Bound}^{parallel})^{2}+( 
R_{Bound-Bound}^{anti-parallel})^{2}
\nonumber \\
&& \ \ \ \ \ \ \ \ \ \ \ \ \ \ \, +
(R_{Free-Free}^{parallel} 
R_{Free-Free}^{anti-parallel})\,\partial^{2}(R_{Free-Free}^{parallel}+ R_
{Free-Free}^{anti-parallel} )
\nonumber \\
&& \ \ \ \ \ \ \ \ \ \ \ \ \ \ \,  +
(R_{Free-Bound}^{parallel} 
R_{Free-Bound}^{anti-
parallel})\,\partial^{2}(R_{Free-Bound}^{parallel}+ R_{Free-Bound}^{anti-parallel} 
)
\nonumber \\
&& \ \ \ \ \ \ \ \ \ \ \ \ \ \ \, 
+
(R_
{Bound-Bound}^{parallel} 
R_{Bound-Bound}^{anti-parallel})\,\partial^{2}(R_{Bound-Bound}^{parallel}+ R_
{Bound-Bound}^{anti-parallel} )
\label{w8},
 \end{eqnarray}
 where   $R_{Bound-Bound}^{anti-parallel}$ is the curvature produced by the interaction 
of 
two 
bound anti-parallel electrons, $R_{Bound-Bound}^{parallel}$ is the curvature created by 
interaction 
of two bound parallel electrons, $R_{Free-Free}^{anti-parallel}$ is the curvature 
produced 
by the 
interaction of two free anti-parallel electrons, $R_{Free-Free}^{parallel}$ is the 
curvature 
created by interaction of two free parallel electrons, $R_{Free-Bound}^{anti-parallel}$ 
is 
the 
curvature produced by the interaction of  free and bound anti-parallel electrons and 
$R_{Free-Free}
^{parallel}$ is the curvature created by interaction of  free and bound parallel 
electrons.

 \section{Defects and $f(R)$  gravity}

We can show that each defect produces a particular form of extended  gravity. 
For example,
 in a graphene without defect where curvature is generated by   parallel spins and neutralized by
curvatures 
of anti-parallel spins, we have the following form of $f(R)$ function:

\begin{eqnarray}
 && \!\!\!\!\!\!\!\!\!\!\!\!\!
f(R)=    \Bigg\{
    2(1-m_{g}^{2})
                \Big[( R_{Free-Free}^{parallel})^{2}+( 
R_{Free-Bound}^{parallel})^{2}+( 
R_{Bound-
Bound}^{parallel})^{2}
\nonumber
\\
&& \ \ \ \ \ \ \ \ \ \ \ \ \ \ \ \ \  \  \ \  + 
(R_{Free-Free}^{parallel} R_{
Free-Free}^{anti-parallel})\,\partial^{2}(R_{Free-Free}^{parallel} )
\nonumber \\
&& \ \ \ \ \ \ \ \ \ \ \ \ \ \ \ \ \  \  \ \ 
+(R_{Free-Bound}^{parallel} 
R_{Free-Bound}^{anti-parallel})\,\partial^{2}(R_{Free-Bound}^{
parallel})
\nonumber \\
&& \ \ \ \ \ \ \ \ \ \ \ \ \ \ \ \ \  \  \ \ 
+
(R_{Bound-Bound}^{parallel} 
R_{Bound-
Bound}^{anti-parallel})\,\partial^{2}(R_{Bound-Bound}^{parallel} ) \Big]
\Bigg\}^{N},
\label{whj10}
    \end{eqnarray}
  where N is the number of atoms in  graphene. 
  
For a graphene structure with heptagonal defects 
where (i.e. the 
curvature produced by parallel spins is larger
  than the curvature produced by anti-parallel spins), some extra terms will be added and 
the form of $f(R)$ gravity is:
 
\begin{eqnarray}
 && \!\!\!\!\!\!\!\!\!\!\!\!\!
 f(R)=
      \Bigg\{
     -(1-m_{g}^{2})
                \Big[( R_{Free-Free}^{parallel})^{2}+( 
R_{Free-Free}^{anti-parallel})^{2}+( 
R_{Free-
Bound}^{parallel})^{2} 
\nonumber 
\\
&& \ \ \ \ \ \ \ \ \ \ \ \ \ \ \ \ \  \  \ \
+
( R_{Free-Bound}^{anti-parallel})^{2}+( 
R_{Bound-Bound}^{
parallel})^{2}+( R_{Bound-Bound}^{anti-parallel})^{2}
\nonumber
\\
&& \ \ \ \ \ \ \ \ \ \ \ \ \ \ \ \ \  \  \ \  + 
(R_{Free-Free}^{parallel} R_{
Free-Free}^{anti-parallel})\,\partial^{2}(R_{Free-Free}^{parallel}+ 
R_{Free-Free}^{anti-parallel} )
\nonumber \\
&& \ \ \ \ \ \ \ \ \ \ \ \ \ \ \ \ \  \  \ \ 
+(R_{Free-Bound}^{parallel} 
R_{Free-Bound}^{anti-parallel})\,\partial^{2}(R_{Free-Bound}^{
parallel}+ R_{Free-Bound}^{anti-parallel} )
\nonumber \\
&& \ \ \ \ \ \ \ \ \ \ \ \ \ \ \ \ \  \  \ \ 
+
(R_{Bound-Bound}^{parallel} 
R_{Bound-
Bound}^{anti-parallel})\,\partial^{2}(R_{Bound-Bound}^{parallel}+ 
R_{Bound-Bound}^{anti-parallel} ) \Big]
\nonumber\\
&& \ \ \ \ \ \ \ \ \ \ \ \ \ \ \ \ \ 
-   m_{g}^{2}\lambda^{2}\delta_{\rho_{1}\sigma_{1}}^{\mu_{1}\nu_{1}}              
\Big[
R^{parallel,\rho_{1}\sigma_{1}}_{Free-Free,\mu_{1}\nu_{1}}+R^{parallel,
\rho_{1}\sigma_{1}}_{Bound-Bound,\mu_{1}\nu_{1}}+R^{parallel,\rho_{1}\sigma_{1}}_{
Free-Bound,\mu_{1}\nu_{1}}
\nonumber\\
&& \ \ \ \ \ \ \ \ \ \ \ \ \ \ \ \ \  \  \ \ 
-
R^{anti-parallel,\rho_{1}\sigma_{1}}_{Free-Free,\mu_{
1}\nu_{1}} -R^{
anti-parallel,\rho_{1}\sigma_{1}}_{Bound-Bound,\mu_{1}\nu_{1}}-R^{anti-parallel,\rho_{1}
\sigma_{1}}_
{Free-
Bound,\mu_{1}\nu_{1}}
\Big]  
\Bigg\}^{N},
\label{wm10}
    \end{eqnarray}
where N is the number of heptagonal molecules. 

For a graphene structure with pentagonal defects  
 (i.e. the 
curvature produced by anti-parallel spins is larger
  than the curvature produced by parallel spins), the sign of $m_{g}^{2}\lambda^{2}$ 
changes:

\begin{eqnarray}
 && \!\!\!\!\!\!\!\!\!\!\!\!\!
 f(R)=
      \Bigg\{
     -(1-m_{g}^{2})
                \Big[( R_{Free-Free}^{parallel})^{2}+( 
R_{Free-Free}^{anti-parallel})^{2}+( 
R_{Free-
Bound}^{parallel})^{2} 
\nonumber 
\\
&& \ \ \ \ \ \ \ \ \ \ \ \ \ \ \ \ \  \  \ \
+
( R_{Free-Bound}^{anti-parallel})^{2}+( 
R_{Bound-Bound}^{
parallel})^{2}+( R_{Bound-Bound}^{anti-parallel})^{2}
\nonumber
\\
&& \ \ \ \ \ \ \ \ \ \ \ \ \ \ \ \ \  \  \ \  + 
(R_{Free-Free}^{parallel} R_{
Free-Free}^{anti-parallel})\,\partial^{2}(R_{Free-Free}^{parallel}+ 
R_{Free-Free}^{anti-parallel} )
\nonumber \\
&& \ \ \ \ \ \ \ \ \ \ \ \ \ \ \ \ \  \  \ \ 
+(R_{Free-Bound}^{parallel} 
R_{Free-Bound}^{anti-parallel})\,\partial^{2}(R_{Free-Bound}^{
parallel}+ R_{Free-Bound}^{anti-parallel} )
\nonumber \\
&& \ \ \ \ \ \ \ \ \ \ \ \ \ \ \ \ \  \  \ \ 
+
(R_{Bound-Bound}^{parallel} 
R_{Bound-
Bound}^{anti-parallel})\,\partial^{2}(R_{Bound-Bound}^{parallel}+ 
R_{Bound-Bound}^{anti-parallel} ) \Big]
\nonumber\\
&& \ \ \ \ \ \ \ \ \ \ \ \ \ \ \ \ \ 
+   m_{g}^{2}\lambda^{2}\delta_{\rho_{1}\sigma_{1}}^{\mu_{1}\nu_{1}}              
\Big[
R^{parallel,\rho_{1}\sigma_{1}}_{Free-Free,\mu_{1}\nu_{1}}+R^{parallel,
\rho_{1}\sigma_{1}}_{Bound-Bound,\mu_{1}\nu_{1}}+R^{parallel,\rho_{1}\sigma_{1}}_{
Free-Bound,\mu_{1}\nu_{1}}
\nonumber\\
&& \ \ \ \ \ \ \ \ \ \ \ \ \ \ \ \ \  \  \ \ 
-
R^{anti-parallel,\rho_{1}\sigma_{1}}_{Free-Free,\mu_{
1}\nu_{1}} -R^{
anti-parallel,\rho_{1}\sigma_{1}}_{Bound-Bound,\mu_{1}\nu_{1}}-R^{anti-parallel,\rho_{1}
\sigma_{1}}_
{Free-
Bound,\mu_{1}\nu_{1}}
\Big]  
\Bigg\}^{N},
\label{wmt10}
    \end{eqnarray}
Here, N is the number of pentagonal molecules. Thus, the gravity which is produced by 
hexagonal 
molecules is very different from gravity produced by pentagonal or heptagonal 
molecules. 
The sign of parameter $m_{g}^{2}\lambda^{2}$ plays a special role in this modelling.

\end{appendix}


\begin{thebibliography}{99}
 

\bibitem{s1}
S. Bera, A. Arnold, F. Evers, R. Narayanan, P. Woelfle,
  {\it{Elastic properties of graphene flakes: boundary effects and lattice vibrations}}
Phys.\ Rev.\ B {\bf 82}, 195445 (2010),
         [\href{http://xxx.lanl.gov/abs/1003.4429}
{{\tt arXiv:1003.4429}}].

\bibitem{Haefner:2014sna} 
  V.~Häfner, J.~Schindler, N.~Weik, T.~Mayer, S.~Balakrishnan, R.~Narayanan, S.~Bera and 
F.~Evers,
  {\it{Density of states in graphene with vacancies: midgap power law and frozen 
multifractality}},
  Phys.\ Rev.\ Lett.\  {\bf 113}, no. 18, 186802 (2014),
       [\href{http://xxx.lanl.gov/abs/1404.6138}
{{\tt arXiv:1404.6138}}].

 

\bibitem{s2a}
P. M. Ostrovsky, I. V. Protopopov, E. J. König, I. V. Gornyi, A. D. Mirlin, M. A. 
Skvortsov,
  {\it{Density of states in a two-dimensional chiral metal with vacancies}},
    Phys.\ Rev.\ Lett.\  {\bf 113}, 186803 (2014),
            [\href{http://xxx.lanl.gov/abs/1404.6139}
{{\tt arXiv:1404.6139}}].

\bibitem{Ulybyshev:2015opa} 
  M.~V.~Ulybyshev and M.~I.~Katsnelson,
  {\it{Magnetism and interaction-induced gap opening in graphene with vacancies or 
hydrogen adatoms: Quantum Monte Carlo study}},
  Phys.\ Rev.\ Lett.\  {\bf 114}, no. 24, 246801 (2015),
         [\href{http://xxx.lanl.gov/abs/1502.01184}
{{\tt arXiv:1502.01184}}].

\bibitem{s2b}
A. A. Stabile, A. Ferreira, J. Li, N. M. R. Peres, J. Zhu, 
  {\it{Electrically tunable resonant scattering in fluorinated bilayer graphene}},
Phys.\ Rev.\ B {\bf 92}, 121411(R) (2015),
         [\href{http://xxx.lanl.gov/abs/1510.01573}
{{\tt arXiv:1510.01573}}].

 

\bibitem{s3}
A. Dasgupta, S. Bera, F. Evers, M. J. van Setten,
  {\it{Quantum Size Effects in the Atomistic Structure of Armchair-Nanoribbons}},
Phys.\ Rev.\ B {\bf 85}, 125433 (2012),
         [\href{http://xxx.lanl.gov/abs/1111.3593}
{{\tt arXiv:1111.3593}}].


 

\bibitem{s4}
 J. Wilhelm, M. Walz, F. Evers,
   {\it{Ab initio quantum transport through armchair graphene nanoribbons: Streamlines in 
the current density}},
Phys.\ Rev.\ B {\bf 89}, 195406 (2014),
         [\href{http://xxx.lanl.gov/abs/1405.3205}
{{\tt arXiv:1405.3205}}].



 
  

\bibitem{Kochetov:2010tm} 
  E.~A.~Kochetov, V.~A.~Osipov and R.~Pincak,
   {\it{Electronic properties of disclinated flexible membrane beyond the inextensional 
limit: Application to graphene}},
  J.\ Phys.\ Condens.\ Matter {\bf 22}, 395502 (2010),
         [\href{http://xxx.lanl.gov/abs/1005.3232}
{{\tt arXiv:1005.3232}}].
 
 
 
\bibitem{s5}
J. Wilhelm, M. Walz, F. Evers,
 {\it{ Ab initio spin-flip conductance of hydrogenated graphene nanoribbons: 
Spin-orbit interaction and scattering with local impurity spins}},
Phys.\ Rev.\ B {\bf 92}, 014405 (2015),
         [\href{http://xxx.lanl.gov/abs/1504.06720}
{{\tt arXiv:1504.06720}}].
 


\bibitem{v5}
Gonzalo J. Olmo, D. Rubiera-Garcia,
 {\it{ Palatini $f(R)$ black holes in nonlinear electrodynamics}},
Phys.\ Rev.\ D {\bf 84}, 124059 (2011),
         [\href{http://xxx.lanl.gov/abs/1110.0850 }
{{\tt arXiv:1110.0850}}].


\bibitem{v6}
Cosimo Bambi, Alejandro Cardenas-Avendano, Gonzalo J. Olmo, D. Rubiera-Garcia,
 {\it{ Wormholes and nonsingular space-times in Palatini f(R)  gravity }},
Phys.\ Rev.\ D {\bf 93}, 064016 (2016),
         [\href{http://xxx.lanl.gov/abs/1511.03755  }
{{\tt  arXiv:1511.03755}}].


\bibitem{v3}
Francisco S. N. Lobo, Gonzalo J. Olmo, D. Rubiera-Garcia,
 {\it{Crystal clear lessons on the microstructure of space-time and modified gravity}},
Phys.\ Rev.\ D {\bf 91}, 124001 (2015),
         [\href{http://xxx.lanl.gov/abs/1412.4499  }
{{\tt  arXiv:1412.4499}}].
 
 
 


 

 \bibitem{B1}
M. A. H. Vozmediano, M. I. Katsnelson, F. Guinea,
   {\it{Gauge fields in graphene}},
  Phys.\ Rept.\  {\bf 496}, 109 (2010),      
  [\href{http://xxx.lanl.gov/abs/1003.5179}
{{\tt arXiv:1003.51790}}].


  
\bibitem{Sepehri:2015ara} 
  A.~Sepehri,
    {\it{Cosmology from quantum potential in brane–anti-brane system}},
  Phys.\ Lett.\ B {\bf 748}, 328 (2015),
    [\href{http://xxx.lanl.gov/abs/1508.01407}
{{\tt arXiv:1508.01407}}].

 
  
   

\bibitem{Sepehri:2016jfx} 
  A.~Sepehri, F.~Rahaman, S.~Capozziello, A.~F.~Ali and A.~Pradhan,
      {\it{Emergence and oscillation of cosmic space by joining M1-branes}},
  Eur.\ Phys.\ J.\ C {\bf 76}, no. 5, 231 (2016),
      [\href{http://xxx.lanl.gov/abs/1604.02451}
{{\tt arXiv:1604.02451}}].
 
 

\bibitem{Sepehri:2014jla} 
  A.~Sepehri, F.~Rahaman, A.~Pradhan and I.~H.~Sardar,
      {\it{Emergence and Expansion of Cosmic Space in BIonic system}},
  Phys.\ Lett.\ B {\bf 741}, 92 (2015),
        [\href{http://xxx.lanl.gov/abs/1501.00428}
{{\tt arXiv:1501.00428}}].
 
  
  
  
 
\bibitem{Sepehri:2015eea} 
  A.~Sepehri, F.~Rahaman, M.~R.~Setare, A.~Pradhan, S.~Capozziello and I.~H.~Sardar,
      {\it{Unifying inflation with late-time acceleration by a BIonic system}},
  Phys.\ Lett.\ B {\bf 747}, 1 (2015),
          [\href{http://xxx.lanl.gov/abs/1505.05105}
{{\tt arXiv:1505.05105}}].
  
  
   

\bibitem{Sepehri:2016sjq} 
  A.~Sepehri,
      {\it{Born–Infeld extension of Lovelock brane gravity in the system of M0-branes and 
its application for the emergence of Pauli exclusion principle in BIonic 
superconductors}},
  Phys.\ Lett.\ A {\bf 380}, 2247 (2016).

  
  
 
\bibitem{Sepehri:2015otm} 
  A.~Sepehri, M.~R.~Setare and S.~Capozziello,
      {\it{Emergence and expansion of cosmic space as due to M0-branes}},
  Eur.\ Phys.\ J.\ C {\bf 75}, no. 12, 618 (2015),
            [\href{http://xxx.lanl.gov/abs/1512.04840}
{{\tt arXiv:1512.04840}}].

\bibitem{Cembranos1} 
  J.A.R. Cembranos, A. Dobado, A.L. Maroto,
  {\it{Brane-world dark matter}},
  Phys.\ Rev.\ Lett.\  {\bf 90}, 241301 (2003),
              [\href{http://xxx.lanl.gov/abs/0302041}
{{\tt arXiv:0302041}}].
 
 \bibitem{Cembranos2} 
   Jose A. R. Cembranos, Antonio L. Maroto,
   {\it{Disformal scalars as dark matter candidates: Branon phenomenology }},
  International Journal of Modern Physics A \  {\bf 31}, 1630015 (2016),
               [\href{http://xxx.lanl.gov/abs/1602.07270 }
 {{\tt arXiv:1602.07270}}].
 
 \bibitem{Cembranos3} 
   J.A.R. Cembranos, A. de la Cruz-Dombriz, V. Gammaldi, A.L. Maroto,
   {\it{Detection of branon dark matter with gamma ray telescopes }},
   Phys.\ Rev.\ D.\  {\bf 85}, 043505 (2012),
               [\href{http://xxx.lanl.gov/abs/1111.4448}
 {{\tt arXiv:1111.4448}}].
 
       
             \bibitem{braneworld1} 
            L. Randall and R. Sundrum,
               {\it{A Large Mass Hierarchy from a Small Extra Dimension }},
               Phys.\ Rev.\ Lett.\  {\bf 83}, 3370 (1999),
                           [\href{http://xxx.lanl.gov/abs/9905221 }
             {{\tt arXiv:9905221}}].
       
 
 
   \bibitem{braneworld2} 
    Gonzalo J. Olmo, D. Rubiera-Garcia,
     {\it{Brane-world and loop cosmology from a gravity-matter coupling perspective }},
     Phys.\ Lett.\ B.\  {\bf 740}, 73 (2015),
                 [\href{http://xxx.lanl.gov/abs/1405.7184 }
   {{\tt arXiv:1405.7184}}].
   
      \bibitem{braneworld3} 
     M.D. Maia, Edmundo M. Monte, J.M.F. Maia,
        {\it{The accelerating universe in brane-world cosmology }},
        Phys.\ Lett.\ B.\  {\bf 585}, 11 (2004),
                    [\href{http://xxx.lanl.gov/abs/0208223 }
      {{\tt arXiv:0208223}}].
      
   
   
\bibitem{Capozziello:2011et} 
  S.~Capozziello and M.~De Laurentis,
  {\it{Extended Theories of Gravity}},
  Phys.\ Rept.\  {\bf 509}, 167 (2011),
              [\href{http://xxx.lanl.gov/abs/1108.6266}
{{\tt arXiv:1108.6266}}].

\bibitem{Capozziello:2007ec}
  S.~Capozziello and M.~Francaviglia,
  {\it Extended Theories of Gravity and their Cosmological and Astrophysical Applications},
  Gen.\ Rel.\ Grav.\  {\bf 40} (2008) 357
 {\tt [arXiv:0706.1146 [astro-ph]].}
  
   
    
  
  

  
\bibitem{Bagger:2007jr} 
  J.~Bagger and N.~Lambert,
  {\it{Gauge symmetry and supersymmetry of multiple M2-branes}},
  Phys.\ Rev.\ D {\bf 77}, 065008 (2008),
          [\href{http://xxx.lanl.gov/abs/0711.0955}
{{\tt arXiv:0711.0955}}].
 
  
   
\bibitem{Myers:1999ps} 
  R.~C.~Myers,
  {\it{Dielectric branes}},
  JHEP {\bf 9912}, 022 (1999),
            [\href{http://xxx.lanl.gov/abs/hep-th/9910053}
{{\tt arXiv:hep-th/9910053}}].
 
  
   

 

\bibitem{Gustavsson:2007vu} 
  A.~Gustavsson,
  {\it{Algebraic structures on parallel M2-branes}},
  Nucl.\ Phys.\ B {\bf 811}, 66 (2009),
              [\href{http://xxx.lanl.gov/abs/0709.1260}
{{\tt arXiv:0709.1260}}].
 
  
    
  


\bibitem{Constable:2001ag} 
  N.~R.~Constable, R.~C.~Myers and O.~Tafjord,
  {\it{NonAbelian brane intersections}},
  JHEP {\bf 0106}, 023 (2001),
                [\href{http://xxx.lanl.gov/abs/hep-th/0102080}
{{\tt arXiv:hep-th/0102080}}].
 
   

 

\bibitem{Sepehri:2016nuv} 
  A.~Sepehri, R.~Pincak and A.~F.~Ali,
  {\it{Emergence of F(R) gravity-analogue due to defects in graphene}},
   [\href{http://xxx.lanl.gov/abs/1606.02039}
{{\tt arXiv:1606.02039}}].
 
    

 
\bibitem{Ho:2008nn} 
  P.~M.~Ho and Y.~Matsuo,
  {\it{M5 from M2}},
  JHEP {\bf 0806}, 105 (2008),
     [\href{http://xxx.lanl.gov/abs/0804.3629}
{{\tt arXiv:0804.3629}}].
 
 
  
  
 
\bibitem{Mukhi:2008ux} 
  S.~Mukhi and C.~Papageorgakis,
  {\it{M2 to D2}},
  JHEP {\bf 0805}, 085 (2008),
       [\href{http://xxx.lanl.gov/abs/0803.3218}
{{\tt arXiv:0803.3218}}].
 
 
  
  
 
\bibitem{Mannheim:2014gba} 
  P.~D.~Mannheim and J.~J.~Poveromo,
  {\it{Gravitational analog of Faraday`s law via torsion and a metric with an 
antisymmetric part}},
  Gen.\ Rel.\ Grav.\  {\bf 46}, no. 10, 1795 (2014),
         [\href{http://xxx.lanl.gov/abs/1406.1470}
{{\tt arXiv:1406.1470}}].
 
  


\bibitem{Ni:2014qfa} 
  W.~T.~Ni,
  {\it{Spacetime structure and asymmetric metric from the premetric formulation of 
electromagnetism}},
  Phys.\ Lett.\ A {\bf 379}, 1297 (2015),
           [\href{http://xxx.lanl.gov/abs/1411.0460}
{{\tt arXiv:1411.0460}}].
 
    

\bibitem{B14a}
R. Pincak, J. Smotlacha, 
  {\it{Analogies in electronic properties of graphene wormhole and perturbed 
nanocylinder}},
  Eur.\ Phys.\ J.\ B {\bf 86},  480 (2013).


\bibitem{Gonzalez:2009je} 
  J.~Gonzalez and J.~Herrero,
  {\it{Graphene wormholes: A Condensed matter illustration of Dirac fermions in curved 
space}},
  Nucl.\ Phys.\ B {\bf 825}, 426 (2010),
             [\href{http://xxx.lanl.gov/abs/0909.3057}
{{\tt arXiv:0909.3057}}].
 
     
  \bibitem{B14b}
   V. Atanasov, A. Saxena,
    {\it{Superconducting tunneling spectroscopy of graphene and graphene nanostructures}},
J.Phys: Cond. Mat. 23 (2011) 175301,
      [\href{http://xxx.lanl.gov/abs/1101.5243}
{{\tt arXiv:1101.5243}}].


\bibitem {exp1}
G. Ozin, A.C. Arsenault,  L. Cademartiri, \textit{Nanochemistry:  A Chemical Approach to 
Nanomateria
ls} 2nd Eds. (Royal Society of Chemistry, 2008) 820p. ISBN 978-1847558954.

\bibitem{exp2}
K.  J. Klabunde,  M. Richards, eds. (2009). \textit {Nanoscale  Materials in Chemistry} 
(2nd ed.). 
Wiley. ISBN 978-0-470-22270-6.

\bibitem{analogue}
M. Visser,  C. Barcel\'o, S. Liberati,  \textit{Analogue models of and for gravity} Gen. 
Rel. Grav. 
 {\bf 34} 1719 (2002).

 


 

  
\end{thebibliography}
\end{document}